\documentclass{article}

\usepackage{times}
\usepackage{amssymb}

\newtheorem{theorem}{Theorem}[section]

\newtheorem{example}[theorem]{Example}
\newtheorem{remark}[theorem]{Remark}

\newtheorem{definition}[theorem]{Definition}
\newtheorem{construction}[theorem]{Construction}

\title{Propositional equality, identity types, \\
and direct computational paths}

\author{Ruy J.G.B.\ de Queiroz \ \ \  Anjolina G.\ de Oliveira\\
Centro de Inform\'atica\\
Universidade Federal de Pernambuco\\
50740-560 Recife, PE, Brazil\\
\{ruy,ago\}@cin.ufpe.br}

\begin{document}

\maketitle

\begin{abstract}
In proof theory the notion of canonical proof is rather basic, and it is usually taken for granted that a canonical proof of a
sentence must be unique up to certain minor syntactical details (such as, e.g., change of bound variables). When setting
up a proof theory for equality one is faced with a rather unexpected situation where there may not be a
unique canonical proof of an equality statement. Indeed, in a (1994--5) proposal for the formalisation of proofs of 
propositional equality in the Curry--Howard style \cite{absldsequality}, we have already uncovered such a peculiarity. Totally independently, and
in a different setting, Hofmann \& Streicher (1994) \cite{hofmann-streicher} have shown how to build a model of Martin-L\"of's Type Theory in
which uniqueness of canonical proofs of identity types does not hold. The intention here is to show that, by considering
as sequences of rewrites and substitution, it comes a rather natural fact that two (or more) distinct proofs may be
yet canonical and are none to be preferred over one another. By looking at proofs of equality as rewriting (or 
computational) paths this approach will be in line with the recently proposed connections between type theory and homotopy theory
via identity types, since elements of identity types will be, concretely, paths (or homotopies).\footnote{The authors would like to thank the anonymous referees for their very careful scrutiny of the paper, leading to significant improvements both in content and presentation. It has to be mentioned that the exchange of e-mail with Thomas Streicher, which happened around June--July 2011 while preparing the first version, was of extraordinary value. Any mistakes or misconceptions, however, are  the fault of the authors of this paper.}

\end{abstract}


\section{Introduction}
There seems to be hardly any doubt that the so-called ``identity types'' are the most intriguing concept of intensional
Martin-L\"of type theory \cite{hofmann-streicher-complete,streicher-uppsala}. From the description of a workshop entitled 
{\em Identity Types -- Topological and Categorical Structure\/}, organised Nov 13--14, 2006, with
support from the Swedish Research Council (VR) and the mathematics departments of Uppsala University and Stockholm University:
\begin{quote}
``The identity type, the type of proof objects for the fundamental propositional
equality, is one of the most intriguing constructions of intensional dependent type
theory (also known as Martin-L\"of type theory). Its complexity became apparent
with the Hofmann-Streicher groupoid model of type theory. This model also
hinted at some possible connections between type theory and homotopy theory
and higher categories. Exploration of this connection is intended to be the main
theme of the workshop.''
\end{quote}
Indeed, a whole new research avenue has recently been explored by people like Vladimir Voevodsky \cite{voevodsky-wollic} and Steve Awodey 
\cite{awodey-survey} in trying to make a bridge between type theory and homotopy theory, mainly via the groupoid structure exposed in the 
Hofmann-Streicher countermodel to the principle of Uniqueness of Identity Proofs (UIP). This has opened the way to, in Awodey's words, 
``a new and surprising connection between Geometry, Algebra, and Logic, which has recently come to light in the form of an interpretation
of the constructive type theory of Per Martin-L\"of into homotopy theory, resulting in new examples of certain algebraic structures which
are important in topology''.

Furthermore, there have been several important  strands in the area of categorical semantics for
Martin-L\"of's type theory, giving rise to rather unexpected links between type theory, abstract homotopy theory and higher-dimensional 
category theory, as pointed out by van den Berg and Garner \cite{berg-garner}. And this is all due to the peculiar structure 
brought about by the so-called identity types:
\begin{quote}
``All of this work can be seen as an elaboration of the following basic idea: that in Martin-L\"of type theory, a type $A$ is analogous to a
topological space; elements $a, b \in A$ to points of that space; and elements of an identity
type $p, q \in {\tt Id}_A(a, b)$ to {\em paths\/} or {\em homotopies\/} $p, q : a \to b$ in $A$.'' \cite{berg-garner}
\end{quote}

\paragraph*{Computational paths.} Motivated by looking at equalities in type theory as arising from the existence of computational paths between two 
formal objects, our purpose here
 is to offer a different perspective on the role and the power of the notion of propositional equality as formalised
in the so-called Curry-Howard functional interpretation. We begin by recalling our previous observation \cite{absldsequality} pertaining to 
the fact that the formulation of the identity type by Martin-L\"of, both in the intensional and in the extensional versions, did not take
into account an important entity, namely, identifiers for sequences of rewrites, and this has led to a false dichotomy. The missing entity has also
made it difficult to formulate the {\em introduction\/} rule for both the intensional and the extensional version without having to resort to the use 
of the reflexivity operator ``${\tt r}$'' as in:
$$\displaystyle{{a:A}\over{{\tt r}(a):{\tt Id}_A(a,a)}}$$
when this should come as a consequence of the general principle of equality saying that, for all elements $a$ of a type $A$, 
equality is by definition a reflexive relation, rather than taking part of the definition of the identity type. 
Instead, if the introduction rule for the identity type takes the form of:
$$\displaystyle{{a=_s b:A} \over {s(a,b):{\tt Id}_A(a,b)}}$$
where the identifier `$s$' is supposed to denote a sequence of rewrites and substitutions which would have started from $a$ and arrived at $b$, it becomes
rather natural to see members of identity types as {\em computational\/} (or {\em rewriting\/}) {\em paths\/}.\footnote{An anonymous reviewer has pointed out that this is common in the ``judgemental" approach to logic: one finds the judgement that underlies a proposition (Cf.\ \cite{PfenningDavies} and \cite{LicataHarper,LicataHarper12}).} By having 
the general rules for equality defined as:
$$\begin{array}{lll}
\mbox{\it reflexivity\/} & \mbox{\it symmetry\/} & \mbox{\it transitivity\/} \\
\  & \  & \ \\
\displaystyle{{x:A} \over {x=_\rho x:A}}\quad &
\displaystyle{{x=_t y:A} \over {y=_{\sigma(t)} x:A}}\quad &
\displaystyle{{x=_t y:A \qquad y=_u z:A} \over {x=_{\tau(t,u)} z:A}}
\end{array}$$
(where `$\sigma$' and `$\tau$' are the symmetry and transitivity rewriting operators) one would then be able to infer that
$$\displaystyle{\displaystyle{{a:A}\over{a=_\rho a:A}}\over{\rho(a):{\tt Id}_A(a,a)}}$$

\paragraph*{Taking an identifier from the meta-language to the object-language.} As we can see from the above example, one may start from `$a:A$', i.e.\ $a$ is
an element of type $A$, and take the $a$ to the object-language by inferring that $\rho(a):{\tt Id}_A(a,a)$. That is to say, in the latter judgement,
the object $a$ is being predicated about in the object language (`$a$ is equal to itself'). It is only via identity types that this can be done in the framework of the Curry-Howard
functional interpretation.

\paragraph*{Iteration.} In the same aforementioned workshop, B.\ van den Berg in his contribution ``Types as weak omega-categories'' draws attention to
the power of the identity type in the iterating types to form a globular set: 
\begin{quote}
Fix a type $X$ in a context $\Gamma$. Define a globular set as follows: $A_0$ consists of the
terms of type $X$ in context $\Gamma$, modulo definitional equality; $A_1$ consists of terms
of the types $Id(X; p;q)$ (in context $\Gamma$) for elements $p, q$ in $A_0$, modulo definitional
equality; $A_2$ consists of terms of well-formed types $Id(Id(X; p;q); r; s)$ (in context
$\Gamma$) for elements $p, q$ in $A_0$, $r, s$ in $A_1$, modulo definitional equality; etcetera...
\end{quote}
Indeed, one may start from:
$$\displaystyle{{p=_r q:X} \over {r(p,q):{\tt Id}_X(p,q)}}\quad \mbox{and} \quad \displaystyle{{p=_s q:X} \over {s(p,q):{\tt Id}_X(p,q)}}$$
and move up one level as in:
$$\displaystyle{{r(p,q)=_u s(p,q):{\tt Id}_X(p,q)} \over {u(r(p,q),s(p,q)):{\tt Id}_{{\tt Id}_X(p,q)}(r(p,q),s(p,q))}}$$
and so on... This has been made precise by a theorem of Peter Lumsdaine
\cite{Lum09} and, independently, by Benno van den Berg and Richard
Garner \cite{BG09,berg-garner} to the effect that, for any type $X$ in Martin-L\"of's (intensional) type theory,  the globular set $\hat{X}$ of terms
of type $X; {\tt Id}_X; {\tt Id}_{{\tt Id}_X};...$ carries a natural weak $\omega$-groupoid structure.

Among other things, this makes it possible to formalise UIP in the theory, as pointed out in Hofmann--Streicher's (1996) survey 
\cite{hofmann-streicher-complete}:
\begin{quote}
``We will call {\em UIP\/} ($U$niqueness of $I$dentity $P$roofs) the following property. If $a_1,a_2$ are objects of type $A$ then for any proofs $p$ and $q$
of the proposition ``$a_1$ {\em equals\/} $a_2$" there is another proof establishing equality of $p$ and $q$. (...) Notice that in traditional logical 
formalism a principle like {\em UIP\/} cannot even be sensibly expressed as proofs cannot be referred to by terms of the object language and thus are not 
within the scope of propositional equality."
\end{quote}
The principle of UIP was originally rendered as \cite{hofmann-streicher}:
$$x: A,\quad p:{\tt Id}_A(x,x)\quad \vdash \quad {\tt Id}_{{\tt Id}_A(x,x)}(p,{\tt r}_A(x))$$
or in the form of a variant for $x$ and $y$ not assumed to be necessarily equal:
$$x: A,\quad y:A,\quad p,q:{\tt Id}_A(x,y)\quad \vdash \quad {\tt Id}_{{\tt Id}_A(x,y)}(p,q)$$
Counter to the principle, put forward by Martin-L\"of, that a type is determined by its canonical object, the model of the identity type 
constructed by Hofmann \& Streicher contains more than one canonical object, and therefore the UIP does not hold. Although this is sharp contrast with
the theory of meaning for type theory as developed mainly by Martin-L\"of, Prawitz and Dummett, it is in perfect agreement with an alternative theory of
meaning based on reduction rules as meaning-giving which we have been advocating for some time now 
\cite{deQueiroz88,deQueiroz89,deQueiroz-phd-thesis,zml1,zml2,deQueiroz91,deQueiroz92,deQueiroz94,deQueiroz2001,deQueiroz2008}.

\paragraph*{Elimination rules and the general principles of equality.} Another aspect of Martin-L\"of's formulation of 
identity types which has posed difficulties in understanding the notion of normal proofs of equality statements is the 
framing of elimination rules for identity types as something of the following sort:
\noindent ${\tt Id}$-{\it elimination\/}
$$\textstyle{{\displaystyle{\ \atop {a:A\quad b:A\quad c:{\tt Id}_A(a,b)}} \quad
\textstyle{{[x:A]} \atop {d(x):C(x,x,{\tt r}(x))}} \quad
\textstyle{{[x:A,y:A,z:{\tt Id}_A(x,y)]} \atop {C(x,y,z)\ type}}} \over
{\displaystyle {\tt J}(c,d):C(a,b,c)}}$$
together with the conversion rule:
\noindent ${\tt Id}$-{\it conversion\/}
$$\displaystyle{{\displaystyle{\ \atop {a:A}}\ 
\displaystyle{{[x:A]} \atop {d(x):C(x,x,{\tt r}(x))}} \ 
\displaystyle{{[x:A,y:A,z:{\tt Id}_A(x,y)]} \atop {C(x,y,z)\ type}}} \over
{{\tt J}({\tt r}(a),d(x))=d(a/x):C(a,a,{\tt r}(a))}}$$

To the elimination operator `${\tt J}$' it is sometimes associated the definition of the usual properties of the 
equality relation:
\begin{quote}
``Surprisingly enough, the ${\tt J}$-eliminator is sufficient for constructing terms refl, symm, trans and subst inhabiting
the types corresponding to the propositions expressing reflexivity, symmetry, transitivity and replacement.'' \cite{hofmann-streicher}
\end{quote}
Same as in \cite{hofmann-streicher-complete}:
\begin{quote}
``The elimination operator $J$ is motivated by the view of $Id(A,\_,\_)$ as an inductively defined family with
with constructor $\mathit{refl}$. Accordingly, $J$ permits one to define an object of type $(a_1,a_2:A)(s:Id(A,a_1,a_2)C(a_1,a_2,s)$
by prescribing its behaviour for arguments of canonical form, i.e.\ $a_1=a_2=a$ and $s=\mathit{refl}(A,a)$.

In the presence of $\Pi$-sets, this elimination operation $J$ allows one to derive the following replacement rule in the
presence of $\Pi$-sets.
$$subst: (A:Set)(P:(a:A)Set)(a_1,a_2:A)(s:Id(a_1,a_2))P(a_1)\to P(a_2)$$
satisfying
$$subst(\mathit{refl}(a),p)=p\mbox{ "}$$
\end{quote}
Nevertheless, in Martin-L\"of's type theory the general properties of equality are given at the level of definitional equality, 
independently of the ${\tt J}$ elimination operator for identity types. Moreover, as soon as the formulation of the rules for the identity types take 
into account the aforementioned ``missing entity'', and thus the existential force of propositional equality, the existence of proofs of transitivity 
and symmetry for propositional equality follow from the application of the rules. Our formulation would be as in:

\medskip

\noindent ${\tt Id}$-{\it elimination\/}
$$\displaystyle{{\ \atop {c:{\tt Id}_A(x,y)}} \quad {{[x=_t y:A]} \atop {d(t):C}}} \over
\displaystyle{{\tt J}(c,\acute{t}d(t)):C}
$$ 
(where $\acute{t}$ is an abstraction over the variable `$t$') with the following conversion rule:

\noindent ${\tt Id}$-{\it conversion\/}
$$\displaystyle{{\displaystyle{{a=_s b:A} \over {s(a,b):{\tt Id}_A(a,b)}}{\tt Id}\mbox{\it -intro\/} \qquad
\displaystyle{{[a=_t b:A]} \atop {d(t):C}}} \over
{{\tt J}(s(a,b),\acute{t}d(t)):C}}{\tt Id}\mbox{\it -elim\/} \qquad
\triangleright_\beta \qquad
\displaystyle{{a=_s b:A} \atop {d(s/t):C}}$$
giving us the equality:
$${\tt J}(s(a,b),\acute{t}d(t))=_\beta d(s/t)$$

With this formulation, we can see that it is by virtue of the elimination rule combined with the general rules of equality on the level of judgements 
that one can prove transitivity and symmetry for propositional equality:
\begin{construction}[${inv}_A$]\ \\
$$\displaystyle{{\displaystyle{{\displaystyle{{\displaystyle{{\displaystyle{\ \atop {[c(x,y):{\tt Id}_A(x,y)]}} \quad
\displaystyle{\displaystyle{{[x=_t y:A]} \over {y=_{\sigma(t)} x:A}} \over {(\sigma(t))(y,x):{\tt Id}_A(y,x)}}} \over {{\tt J}(c(x,y),(\sigma(\acute{t}))(y,x)):{\tt Id}_A(y,x)}}}
\over {\lambda c.{\tt J}(c(x,y),(\sigma(\acute{t}))(y,x)):{\tt Id}_A(x,y)\to{\tt Id}_A(y,x)}}}
\over {\lambda y.\lambda c.{\tt J}(c(x,y),(\sigma(\acute{t}))(y,x)):\Pi y:A.({\tt Id}_A(x,y)\to{\tt Id}_A(y,x))}}}
\over {\lambda x.\lambda y.\lambda c.{\tt J}(c(x,y),(\sigma(\acute{t}))(y,x)):\Pi x:A.\Pi y:A.({\tt Id}_A(x,y)\to{\tt Id}_A(y,x))}}
$$
where $\sigma$ is the symmetry operator introduced by the general rule of symmetry given as part of the definition of
equality on the lefthand side. 
\end{construction}

\begin{construction}[${cmp}_A$]\ \\
$${{{{{{{{\ \atop {[w(x,y):{\tt Id}_A(x,y)]}} \quad {{\ \atop {[s(y,z):{\tt Id}_A(y,z)]}}\quad {{{[x=_t y:A] \quad [y=_u z:A]} \over {x=_{\tau(t,u)} z:A}} \over {(\tau(t,u))(x,z):{\tt Id}_A(x,z)}} \over {{\tt J}(s(y,z),\acute{u}(\tau(t,u))(x,z)):{\tt Id}_A(x,z)}}
\over {{\tt J}(w(x,y),\acute{t}{\tt J}(s(y,z),\acute{u}(\tau(t,u))(x,z))):{\tt Id}_A(x,z)}}}
\over {\lambda s.{\tt J}(w(x,y),\acute{t}{\tt J}(s(y,z),\acute{u}(\tau(t,u))(x,z))):({\tt Id}_A(y,z)\to{\tt Id}_A(x,z))}}
\over {\lambda w.\lambda s.{\tt J}(w(x,y),\acute{t}{\tt J}(s(y,z),\acute{u}(\tau(t,u))(x,z))):({\tt Id}_A(x,y)\to({\tt Id}_A(y,z)\to{\tt Id}_A(x,z)))}}
\over {\lambda z.\lambda w.\lambda s.{\tt J}(w(x,y),\acute{t}{\tt J}(s(y,z),\acute{u}(\tau(t,u))(x,z))):\Pi z:A.({\tt Id}_A(x,y)\to({\tt Id}_A(y,z)\to{\tt Id}_A(x,z)))}}
\over {\lambda y.\lambda z.\lambda w.\lambda s.{\tt J}(w(x,y),\acute{t}{\tt J}(s(y,z),\acute{u}(\tau(t,u))(x,z))):\Pi y:A.\Pi z:A.({\tt Id}_A(x,y)\to({\tt Id}_A(y,z)\to{\tt Id}_A(x,z)))}}
\over {\lambda x.\lambda y.\lambda z.\lambda w.\lambda s.{\tt J}(w(x,y),\acute{t}{\tt J}(s(y,z),\acute{u}(\tau(t,u))(x,z))):\Pi x:A.\Pi y:A.\Pi z:A.({\tt Id}_A(x,y)\to({\tt Id}_A(y,z)\to{\tt Id}_A(x,z)))}}
\atop \ 
$$
\end{construction}
The final proof terms above are called, respectively, ${inv}_A$ and ${cmp}_A$ by Streicher \cite{streicher-uppsala}:
\begin{quote}
``Using ${\tt J}$ one can define operations

${cmp}_A \in (\Pi x, y, z:A) {\tt Id}_A(x, y)\to{\tt Id}_A(y, z)\to {\tt Id}_A(x, z)$

${inv}_A \in (\Pi x, y:A) {\tt Id}_A(x, y) \to {\tt Id}_A(y, x)$ ''
\end{quote}

\paragraph*{The groupoid laws.} It so happens that the existence of ${cmp}_A$ and ${inv}_A$ validates the following 
groupoid laws as pointed out in Streicher's talk at the aforementioned workshop ``Identity Types vs.\ Weak $\omega$-Groupoids --
Some Ideas and Problems" \cite{streicher-uppsala}:
$$\begin{array}{ll}
(a) & (\Pi x, y, z, u:A)\\
& (\Pi f:{\tt Id}_A(x, y))(\Pi g:{\tt Id}_A(y, z))(\Pi h:{\tt Id}_A(z, u))\\
& {\tt Id}_{{\tt Id}_A(x,u)}({cmp}_A(f, {cmp}_A(g, h)), {cmp}_A({cmp}_A(f, g), h))\\
(b) & (\Pi x, y:A) {\tt Id}_{{\tt Id}_A(x,y)}({cmp}_A({\tt r}(x), f), f) \land {\tt Id}_{{\tt Id}_A(y,x)}({cmp}_A(g, {\tt r}(y)), g)\\
(c) & (\Pi x, y:A)(\Pi f:{\tt Id}_A(x, y))\\
& {\tt Id}_{{\tt Id}_A(x,x)}({cmp}_A(f, {inv}_A(f)), {\tt r}(x)) \land {\tt Id}_{{\tt Id}_A(y,y)}({cmp}_A({inv}_A(f), f), {\tt r}(y))
\end{array}$$
This makes type $A$ an internal groupoid where the groupoid equations hold only in the sense of propositional equality.
Indeed, via the reduction rules defined over the terms corresponding to equality proofs, one can see that the laws are 
validated. Just to motivate the reductions between proofs of equality, let us recall that
the rule of {\em symmetry\/} is the only rule which changes the direction of an equation.
So, its use must be controlled. Here we give two reductions over proofs of equality which are related to such a 
need for controlling the use of symmetry. (The rewriting system with all reductions between terms of identity types is given in Definition~\ref{rewriting-system}.)

\begin{definition}[reductions involving $\rho$ and $\sigma$]
$$ \displaystyle{ x=_\rho x : {A} \over x =_{\sigma(\rho)} x : {A}}  \quad \triangleright_{sr} \quad  x =_\rho x : {A}$$
$$ \displaystyle{\displaystyle{ x=_r y : {A} \over y =_{\sigma(r)
} x : {A}} \over x =_{\sigma(\sigma(r)) }y:A}  \quad \triangleright_{ss} \quad
 x=_r y : {A}$$
Associated rewritings:\\
$\sigma(\rho)\triangleright_{sr} \rho$ \\
$\sigma(\sigma(r))\triangleright_{ss} r$ 
\end{definition}
By applying the rule of propositional equality to the level of ${\tt Id}_{{\tt Id}_A(x,x)}$ we can get:
$$\displaystyle{{\displaystyle{ x=_\rho x : {A} \over x =_{\sigma(\rho)} x : {A}}} \over {(\sigma(\rho))(x):{\tt Id}_A(x,x)}}  \quad \triangleright_{sr} \quad  \displaystyle{x =_\rho x : {A}\over \rho(x):{\tt Id}_A(x,x)}$$
i.e., `$(\sigma(\rho))(x)$' and `$\rho(x)$' are two equal proofs of ${\tt Id}_A(x,x)$. So,
$$\displaystyle{{\sigma(\rho)=_{sr} \rho:{\tt Id}_A(x,x)} \over {(sr)(\sigma(\rho),\rho):{\tt Id}_{{\tt Id}_A(x,x)}}(\sigma(\rho),\rho)}$$ 
And similarly:
$$ \displaystyle{{\displaystyle{\displaystyle{ x=_r y : {A} \over y =_{\sigma(r)
} x : {A}} \over x =_{\sigma(\sigma(r)) }y:A}} \over {(\sigma(\sigma(r)))(x,y):{\tt Id}_A(x,y)}}  \quad \triangleright_{ss} \quad
 \displaystyle{x=_r y : {A}\over {r(x,y):{\tt Id}_A(x,y)}}$$
Thus:
$$\displaystyle{{\sigma(\sigma(r))=_{ss} r:{\tt Id}_A(x,y)} \over {(ss)(\sigma(\sigma(r)),r):{\tt Id}_{{\tt Id}_A(x,y)}}(\sigma(\sigma(r)),r)}$$ 

Similarly, the transitivity operation on proofs of equality brings us the following reductions:

\begin{definition}[$\tau$ and $\tau$]\ \\
$\displaystyle{\displaystyle{ x=_t y:A \quad y =_r w: {A} \over x =_{\tau(t,r)}w: {A}} \quad
\displaystyle{ \atop w =_s z : {A}} \over x =_{\tau (\tau (t,r) , s)} z : {A}}$

\hfill{$\triangleright_{tt} \quad  
\displaystyle{\displaystyle{ \atop x =_t y : {A}} \quad  
\displaystyle{ y=_r w : {A} \quad w=_s z : {A} \over  y=_{\tau(r,s)} z : {A}} \over x=_{\tau(t,\tau(r,s))} z :  
{A}}$}

\smallskip\noindent
Associated rewriting: \\
$\tau(\tau (t,r),s) \triangleright_{tt} \tau(t,\tau(r,s))$
\end{definition}
So, 

\noindent $\displaystyle{\displaystyle{\displaystyle{ x=_t y:A \quad y =_r w: {A} \over x =_{\tau(t,r)}w: {A}} \quad
\displaystyle{ \atop w =_s z : {A}} \over x =_{\tau (\tau (t,r) , s)} z : {A}} \over 
(\tau(\tau(t,r),s))(x,z):{\tt Id}_A(x,z)}$

\hfill{$\triangleright_{tt} \quad  
\displaystyle{\displaystyle{\displaystyle{ \atop x =_t y : {A}} \quad  
\displaystyle{ y=_r w : {A} \quad w=_s z : {A} \over  y=_{\tau(r,s)} z : {A}} \over x=_{\tau(t,\tau(r,s))} z :  
{A}} \over
(\tau(t,\tau(r,s)))(x,z):{\tt Id}_A(x,z)}$}

Thus
$$\displaystyle{{\tau(\tau(t,r),s)=_{tt} \tau(t,\tau(r,s)):{\tt Id}_A(x,z)} \over {(tt)(\tau(\tau(t,r),s),\tau(t,\tau(r,s))):{\tt Id}_{{\tt Id}_A(x,z)}}(\tau(\tau(t,r),s),\tau(t,\tau(r,s)))}$$

Notice that, although the type ${\tt Id}_{{\tt Id}_A(x,z)}(\tau(\tau(t,r),s),\tau(t,\tau(r,s)))$ is inhabited, i.e.\ there is a proof-term of that type,
this does not pressupose that, seeing $t,r,s$ as functions, $t\circ(r\circ s)=(t\circ r)\circ s$. 
This is similar to Hofmann-Streicher's statement on Proposition
4.1 of \cite{hofmann-streicher-complete}:
\begin{quote}
``If $a_1,a_2,a_3,a_4:A$ and $s_1:{\tt Id}_A(a_1,a_2)$ and $s_2:{\tt Id}_A(a_2,a_3)$ and $s_3:{\tt Id}_A(a_3,a_4)$ then
$$trans(s_3,(trans(s_2,s_1))=_{prop}trans(trans(s_3,s_2),s_1)\mbox{"}$$
\end{quote}
(where `$s_1=_{prop}s_2$' meant that the type ${\tt Id}_{{\tt Id}_A(a_1,a_2)}(s_1,s_2)$ was inhabited). Notice again that it was not required 
that $s_3\circ(s_2\circ s_1)=(s_3\circ s_2)\circ s_1$.

The same observation is made by Warren \cite{warren-thesis}:
\begin{quote}
``For example, given terms $f$ of type ${\tt Id}_A(a, b)$
and $g$ of type ${\tt Id}_A(b, c)$, there exists a ``composite" $(g \cdot f)$ of type ${\tt Id}_A(a, c)$. However,
this composition and the identities mentioned above fail to satisfy the actual
category axioms ``on-the-nose", but only up to the existence of terms of further
``higher-dimensional" identity types. Thus, given $f$ and $g$ as above together with a
further term $h$ of type ${\tt Id}_A(c, d)$, the type
$${\tt Id}_{{\tt Id}_A(a,d)}(h \cdot (g \cdot f), (h \cdot g) \cdot f)$$
is inhabited; but it is not in general the case that $h\cdot(g\cdot f) = (h\cdot g)\cdot f$."
\end{quote}  
The fact that the structure brought about by identity types satisfy the groupoid laws, but only `at the propositional equality', is also highlighted
by Steve Awodey in his recent survey:
\begin{quote}
``In the intensional theory, each type $A$ is thus endowed by the identity types ${\tt Id}_A(a, b)$ with a non-trivial structure. Indeed, this structure
was observed by Hofmann and Streicher in [HS98] to satisfy conditions analogous to the familiar laws for groupoids. Specifically, the
posited refexivity of propositional equality produces identity proofs ${\tt r}(a) : {\tt Id}_A(a, a)$ for any term $a : A$, playing the
 role of a unit arrow $1_a$ for $a$; and when $f : {\tt Id}_A(a, b)$ is an identity proof, then (corresponding
to the symmetry of identity) there also exists a proof $f^{-1} : {\tt Id}_A(b, a)$, to be thought of as the inverse of $f$; finally, when $f : {\tt Id}_A(a, b)$ and $g : {\tt Id}_A(b, c)$ are identity proofs, then (corresponding to transitivity)
there is a new proof $g \circ f : {\tt Id}_A(a, c)$, thought of as the composite of $f$ and $g$.
 Moreover, this structure on each type $A$ can be shown to satisfy the usual groupoid laws, but significantly, only {\bf up to propositional
equality}." \cite{awodey-survey}
\end{quote}

In what follows we will spell out a refinement of the approach to propositional equality which was presented in a previous paper on the functional
interpretation of direct computations \cite{direct-computations}. The intention, as already put forward above, is to offer a formulation of a proof theory
for propositional equality very much in the style of identity types which, besides being a reformulation of Martin-L\"of's own intensional identity types
into one which dissolves what we see as a false dichotomy, turns out to validate the groupoid laws as uncovered by Hofmann \& Streicher as well as to refute
the principle of uniqueness of identity proofs. So, we are left with a sort of `weak' type theory in which the connections between the deductive system and the semantics of ``a type is a space of paths", which is not so obvious in Martin-L\"of's own formulation of intensional type theory, becomes rather natural: propositional equality is indeed the type of (computational) sequences/paths between two elements of a type.

The main point of this paper is to show the connections between the approach to propositional equality that we have been developing since the early 1990's to the one put forward by the Hofmann-Streicher-Voevodsky-Awodey tradition which has now come to be documented in a phenomenal and collective production entitled {\em Homotopy Type Theory\/} \cite{HoTT} (Aug 2013). This is useful because, in spite of the differences in details, the approaches seem to have arrived at similar conclusions: elements of the identity type are paths/sequences-of-rewrites from an object to another object of a certain type, which gives rise to all these exciting connections to homotopy. The aim is not to make a formal comparison of the two approaches, but rather to explore both the similarities and the differences between them, and at the same time expose the interesting convergence of groundbreaking conclusions with respect to the connections between type theory and homotopy theory. 

\section{Normal form for proofs of equality}

The clarification of the notion of normal form for equality reasoning took
an important step with the work of Statman in the late
1970's \cite{statman77,statman78}. The concept of {\em direct computation\/}
was instrumental in the development of Statman's approach.
By way of motivation, let us take a simple
example from the $\lambda$-calculus.
$$\begin{array}{lllllll}
(\lambda x.(\lambda y.yx)(\lambda w.zw))v & \triangleright_\eta & (\lambda x.(\lambda y.yx)z)v & \triangleright_\beta &  (\lambda y.yv)z & \triangleright_\beta & zv\\
(\lambda x.(\lambda y.yx)(\lambda w.zw))v & \triangleright_\beta & (\lambda x.(\lambda w.zw)x)v & \triangleright_\eta & (\lambda x.zx)v & \triangleright_\beta & zv\\
(\lambda x.(\lambda y.yx)(\lambda w.zw))v & \triangleright_\beta & (\lambda x.(\lambda w.zw)x)v & \triangleright_\beta & (\lambda w.zw)v & \triangleright_\eta & zv
\end{array}$$
There is at least one sequence of conversions, i.e.\ one computational path, from the initial term to the
final term. (In this case we have given three!) Thus, in the formal theory of
$\lambda$-calculus, the term $(\lambda x.(\lambda y.yx)(\lambda w.zw))v$ is
declared to be {\bf equal} to $zv$.

Now, some natural questions arise:
\begin{enumerate}
\item Are the sequences/paths themselves
{\em normal\/}?
\item Are there non-normal sequences/paths?
\item If yes, how are the latter to
be identified and (possibly) normalized?
\item What happens if general rules of equality are involved? 
\end{enumerate}

Of course, if one considers only the $\beta$-contractions, the traditional choice is for the so-called
{\em outermost and leftmost reduction\/} \cite{hindleyseldin}.

Nevertheless, we are interested in an approach to these questions that would be 
applicable both to $\lambda$-calculus and to proofs in Gentzen's style Natural Deduction.
As rightly pointed out by Le~Chenadec in \cite{chenadec89},
the notion of normal proof has been somewhat neglected by the systems of
equational logic:
``In proof-theory, since the original work of Gentzen (1969) on sequent
calculus, much work has been devoted to the normalization process of
various logics, Prawitz (1965), Girard (1988). Such an analysis was
lacking in equational logic (the only exceptions we are aware of are
Statman (1977), Kreisel and Tait (1961)).''
The works of Statman \cite{statman77,statman78}
and Le~Chenadec \cite{chenadec89} represent significant attempts to fill
this gap.
Statman studies proof transformations for the equational calculus $E$
of Kreisel--Tait \cite{kretait61}.  Le~Chenadec
defines an equational proof system (the {\em LE system})
and gives a normalization procedure.

\paragraph*{What is a proof of an equality statement?} 
The so-called Brouwer-Heyting-Kolmogorov Interpretation defines logical connectives by
taking proof, rather than truth-values, as a primitive notion:
\begin{tabbing}
\textbf{a proof of the proposition:} \hbox{\ \ \ \ \ } \= \textbf{is given by:} \\
$A\land B$ \> a proof of $A$ \textbf{and} a proof of $B$ \\ 
$A\lor B$ \> a proof of $A$ \textbf{or} a proof of $B$ \\
$A\rightarrow B$ \> a \textbf{function} that turns a proof of $A$ into a proof of $B$ \\
$\forall x^D.P(x)$ \> a \textbf{function} that turns an element $a$ into a proof of $P(a)$ \\
$\exists x^D.P(x)$ \> an element $a$ (witness) \textbf{and a proof of} $P(a)$ \\
\end{tabbing}
Based on the Curry-Howard functional interpretation of logical connectives, one can
formulate the BHK-interpretation in formal terms as following:
\begin{tabbing}
\textbf{a proof of the proposition:} \hbox{\ \ \ \ \ } \= \textbf{has the canonical form of:} \\
$A\land B$ \> $\langle p,q\rangle$ where $p$ is a proof of $A$ and $q$ is a proof of $B$ \\
$A\lor B$ \> $i(p)$ where $p$ is a proof of $A$ or $j(q)$ where $q$ is a proof of $B$ \\
\> (`$i$' and `$j$' abbreviate `into the left/right disjunct') \\
$A\rightarrow B$ \> $\lambda x.b(x)$ where $b(p)$ is a proof of B \\
\> provided $p$ is a proof of A \\
$\forall x^A.B(x)$ \> $\Lambda x.f(x)$ where $f(a)$ is a proof of $B(a)$ \\ 
\> provided $a$ is an arbitrary individual chosen\\
\> from the domain $A$\\ 
$\exists x^A.B(x)$ \> $\varepsilon x.(f(x),a)$ where $a$ is a witness\\
\> from the domain $A$, $f(a)$ is a proof of $B(a)$ \\
\end{tabbing}
(The term `$\varepsilon x.(f(x),a)$' is framed so as to formalise the notion of
a function carrying its own argument \cite{ldsexistential}.)

A question remains, however:
\begin{quote}
What is a proof of an equality statement?
\end{quote}
An answer to such a question will help us extend the BHK-interpretation with an explanation of
what is a proof of an equality statement:
\begin{tabbing}
\textbf{a proof of the proposition:} \hbox{\ \ \ \ \ } \= \textbf{is given by:} \\
\\
$t_1= t_2$ \> ? \\
\> (Perhaps a sequence of rewrites \\
\> starting from $t_1$ and ending in $t_2$?) \\
\end{tabbing}
Two related questions naturally arise:
\begin{enumerate}
\item What is the logical status of the symbol ``$=$''?

\item What would be a canonical/direct proof of $t_1=t_2$?
\end{enumerate}

In a previous work \cite{NDchapter} we have tried to show how the framework of
labelled natural deduction
can help us formulate a proof theory for the ``logical connective'' of
propositional equality.\footnote{An old question is in order here: what is a
logical connective? We shall take it that from the point of view of proof
theory (natural deduction style) a
logical connective is whatever logical symbol which is analysable into rules of
{\em introduction\/} and {\em elimination\/}.}
The connective is meant to be used in reasoning about
equality between referents (i.e.\ the terms alongside formulas/types), as
well as with a general notion of substitution which is needed for the
characterization of the so-called {\em term declaration logics\/}
\cite{Aczel91}.

In order to account for the distinction between the equalities that are:
\begin{enumerate}
\item[] {\em definitional\/}, i.e.\ those equalities that are given as rewrite
rules (equations), orelse originate from general functional principles (e.g.\ 
$\beta$, $\eta$, etc.),
\end{enumerate}
and those that are:
\begin{enumerate}
\item[] {\em propositional\/}, i.e.\ the equalities that are supported (or
otherwise) by an evidence (a composition of rewrites),
\end{enumerate}
we need to provide for an equality sign as a symbol for {\em rewrite\/}
(i.e.\ as part of the functional calculus on the terms),
and an equality sign as a symbol for a {\em relation\/} between referents
(i.e.\ as part of the logical calculus on the formulas/types).

Single steps of reduction come from definitional equalities, and those single steps can be composed leading to sequences of rewrites, which can then turned into a propositional equality. It helps to remember that in ``$t:A$", the logical interpretation is that ``$t$" is a (functional) term, and ``$A$" is a statement. So, the equality is propositional when it is a statement, i.e., in ``$q:{\tt Id}$", ``${\tt Id}$" is a statement which is supported by the term ``$q$" (which, in its turn, can be an equational term like ``$a=_s b$"). So, while ``$q$" will carry definitional content (be it single or composed), ``${\tt Id}$" will carry propositional content.

\paragraph*{Definitional equalities.} Let us recall from the theory of
$\lambda$-calculus, that:

\begin{definition}[\cite{hindleyseldin}, (Definition 6.2 and Notation 7.1)] The formal theory of $\lambda\beta\eta$
equality has the following axioms:
$$\begin{array}{ll}
(\alpha) & \lambda x.M=\lambda y.[y/x]M \qquad\quad (y\notin FV(M))\\
(\beta) & (\lambda x.M)N = [N/x]M\\
(\eta) & (\lambda x.Mx) = M \qquad\qquad\quad (x\notin FV(M))\\
(\rho) & M=M\\
\end{array}$$
and the following inference rules:
$$\begin{array}{llll}
(\mu) & \displaystyle{{M=M'}\over {NM=NM'}}\qquad & (\tau) & \displaystyle{{M=N \qquad N=P} \over {M=P}}\\
&\\
(\nu) & \displaystyle{{M=M'}\over {MN=M'N}}\qquad & (\sigma) & \displaystyle{{M=N} \over {N=M}} \\
&\\
(\xi) & \displaystyle{{M=M'}\over {\lambda x.M=\lambda x.M'}}\\
&\\
(\zeta) & \displaystyle{{Mx=Nx}\over {M=N}} && \mbox{if } x\notin FV(MN)
\end{array}$$
\end{definition}
In Martin-L\"of's type theory the axioms are introduced as:
$$\begin{array}{llll}
(\beta) & \displaystyle{\displaystyle{{\ \atop {N:A}}\quad \displaystyle{{[x:A]} \atop {M:B}}}\over (\lambda x.M)N = M[N/x]:B}\\
&\\
(\eta) & \displaystyle{{M:(\Pi x:A)B}\over (\lambda x.Mx) = M:(\Pi x:A) B} \ (x\notin FV(M))\\
&\\
(\rho) & \displaystyle{M:A\over M=M:A}\\
&\\
(\mu) & \displaystyle{{M=M':A\quad N:(\Pi x:A)B}\over {NM=NM':B}}\qquad & (\tau) & \displaystyle{{M=N:A \qquad N=P:A} \over {M=P:A}}\\
&\\
(\nu) & \displaystyle{{N:A\quad M=M':(\Pi x:A)B}\over {MN=M'N:B}}\qquad & (\sigma) & \displaystyle{{M=N:A} \over {N=M:A}} \\
&\\
(\xi) & \displaystyle{\displaystyle{{[x:A]} \atop {M=M':B}}\over {\lambda x.M=\lambda x.M':(\Pi x:A)B}}
\end{array}$$

\paragraph*{Propositional equality.} Again, let us recall from the formal theory of
$\lambda$-calculus, that:

\begin{quote}
{\bf Definition 1.37 ($\beta$-equality)} \cite{hindleyseldin} \\
We say that $P$ is $\beta$-equal or $\beta$-convertible to $Q$ (notation $P=_\beta Q$)
iff $Q$ can be obtained from $P$ by a finite (perhaps empty) series of $\beta$-contractions
and reversed $\beta$-contractions and changes of bound variables. That is,
$P=_\beta Q$ iff there exist $P_0,\ldots,P_n$ ($n\geq 0$) such that
$$\begin{array}{c}
(\forall i\leq n-1)(P_i\triangleright_{1\beta}P_{i+1}\mbox{  or  }P_{i+1}\triangleright_{1\beta}P_i\mbox{  or  } P_i\equiv_\alpha P_{i+1}).\\
P_0\equiv P, \qquad P_n\equiv Q.
\end{array}$$
\end{quote}
NB: equality with an {\bf existential} force.

The same happens with $\lambda\beta\eta$-equality:
\begin{quote}
{\bf Definition 7.5 ($\lambda\beta\eta$-equality)} \cite{hindleyseldin} \\
The equality-relation determined by the theory $\lambda\beta\eta$ is called
$=_{\beta\eta}$; that is, we define
$$M=_{\beta\eta}N\quad\Leftrightarrow\quad\lambda\beta\eta\vdash M=N.$$
\end{quote}
Note again that two terms are $\lambda\beta\eta$-equal if {\bf there exists}
a proof of their equality in the theory of $\lambda\beta\eta$-equality.

\begin{remark}
In setting up a set of Gentzen's ND-style rules for equality we need to account for:\\
1. the dichotomy definitional versus propositional equality;\\
2. there may be more than one {\em normal\/} proof of a certain equality statement;\\
3. given a (possibly non-normal) proof, the process of bringing it to a normal
form should be finite and confluent.
\end{remark}

\paragraph*{The missing entity.} Within the framework of the functional interpretation ({\em \`a la\/}
Curry--Howard \cite{howard80}), the definitional equality is often considered by reference to a
judgement of the form:
$$a = b:A$$
which says that $a$ and $b$ are equal elements from domain or type $A$. Notice that
the `reason' why they are equal does not play any part in the judgement. This
aspect of `forgetting contextual information' is, one might say, the first step
towards `extensionality' of equality, for whenever one wants to introduce
intensionality into a logical system one invariably needs to introduce
information of a `contextual' nature, such as, where the identification of two
terms (i.e.\ equation) comes from.

We feel that a first step towards finding an alternative formulation of the
proof theory for propositional equality which takes care of the intensional
aspect is to allow the `reason' for the equality to play a more significant part
in the form of judgement. We also believe that from the point of view of the
logical calculus, if there is a `reason' for two expressions to be considered
equal, the proposition asserting their equality will be true, regardless of what
particular composition of rewrites (definitional equalities) amounts to the
evidence in support of the proposition concerned. Given these general
guidelines, we shall provide what may be seen as a middle ground solution
between the intensional \cite{Martin-Lof75:itt, Martin-Lof75:models} and the extensional
\cite{Martin-Lof82:itt} accounts of Martin-L\"of's propositional equality.
The intensionality is taken care by the functional calculus on the labels,
while the extensionality is catered by the logical calculus on the formulas.
In order to account for the intensionality in the labels, we shall make the
composition of rewrites (definitional equalities) appear as indexes of the
equality sign in the judgement with a variable denoting a sequence of equality
identifiers (we have seen that in the Curry--Howard functional interpretation there
are at least four `natural' equality identifiers: $\beta$, $\eta$, $\xi$ and
$\mu$). So,
instead of the form above, we shall have the following pattern for the equality
judgement:
$$a =_s b:A$$
where `$s$' is meant to be a sequence of equality identifiers.

In the sequel we shall be discussing in some detail the need to identify the
kind of definitional equality, as well as the need to have a logical connective
of `propositional equality' in order to be able to reason about the functional
objects (those to the left hand side of the `:' sign).

\paragraph*{Term rewriting.}
Deductive systems based on the  Curry--Howard isomorphism
\cite{howard80} have an interesting feature: normalization and strong
normalization (Church--Rosser property) theorems can be proved by reductions on
the terms of the functional calculus. Exploring this important characteristic,
we have proved these theorems for the {\em Labelled Natural Deduction\/} --
{\em LND\/} \cite{ldsintro,lndlivro} via a term rewriting system
constructed from the {\em LND\/}-terms of the functional calculus
\cite{deOliveira2}.
Applying this same technique to the {\em LND\/}
 equational fragment, we obtain the normalization theorems for the equational
 logic of the {\em Labelled Natural Deduction\/} System
\cite{teseju,deOliveira3,deOliveira4}.

This technique is used given the possibility of defining two measures of redundancy  for the
 {\em LND\/} system that can be dealt with in the object level: the terms on the
 functional calculus and the {\em rewrite reason\/} (composition of rewrites), 
the latter being indexes of the equations in the {\em LND\/} equational fragment.

In the {\em LND\/} equational logic \cite{absldsequality}, the equations have the following 
pattern:

$$ a=_s b : A$$
where one is to read: $a$ is equal to  $b$ {\em because\/} of `$s$' (`$s$'
being the {\em rewrite reason\/}); `$s$' is a term denoting a sequence of equality 
identifiers ($\beta$, $\eta$, $\alpha$, etc.), i.e.\ a composition of 
rewrites. In other words, `$s$' denotes the 
{\em computational path\/} from $a$ to $b$.

In this way, the {\em rewrite reason\/} (reason, for short) represents an
 {\bf orthogonal measure of redundancy} for the {\em LND\/}, which makes the
  {\em LND} equational fragment an ``enriched'' system of equational logic. 
Unlike the traditional equational logic systems, in {\em LND\/} equational
 fragment there is a gain in local control by the use of {\em reason\/}. 
All the proof steps are recorded in the composition of rewrites (reasons). Thus,
 consulting the reasons, one should be able to see whether the proof has the
 normal form. We have then used this powerful mechanism of controlling
 proofs to present a precise normalization procedure for the {\em LND\/}
 equational fragment. Since the reasons can be dealt with in the object level,
 we can employ a computational method to prove the normalization theorems:
 we built a term rewriting system based on an algebraic calculus on
 the ``{\em rewrite reasons}'', which compute normal proofs. With this we
believe we are making a step towards filling a gap in the literature on
equational logic and on proof theory (natural deduction).

\paragraph*{Kreisel--Tait's system.}
In \cite{kretait61} Kreisel and Tait define the system $E$ for equality
reasoning as consisting of axioms of the form $t=t$, and the following
rules of inference:
$$\begin{array}{lc}
(E1) & \displaystyle{{E[t/x] \quad t=u}\over {E[u/x]}}\\
&\\
(E2) & \displaystyle{{s(t)=s(u)}\over {t=u}}\\
&\\
(E3) & \displaystyle{{0=s(t)}\over A}\quad\mbox{for any formula }A\\
&\\
(E4_n) & \displaystyle{{t=s^n(t)}\over A}\quad\mbox{for any formula }A
\end{array}$$
where $t$ and $u$ are terms, `$0$' is the first natural number (zero),
`$s($-$)$' is the successor function.

\paragraph*{Statman's normal form theorem.}
In order to prove the normalization results for the calculus $E$ Statman
defines two subsets of $E$: (i) a natural deduction based calculus for
equality reasoning $NE$; (ii) a sequent style calculus $SE$.

The $NE$ calculus is defined as having axioms of the form $a=a$, and the
rule of substituting equals for equals:
$$\begin{array}{lc}
(=) & \displaystyle{{E[a/u] \quad a\approx b}\over {E[b/u]}}
\end{array}$$
where $E$ is any set of equations, and $a\approx b$ is ambiguously $a=b$
and $b=a$.

Statman arrives at various important results on normal forms and bounds for
proof search in $NE$. In this case, however, a rather different notion of
normal form is being used: the `cuts' do not arise out of an {\em inversion
principle\/}, as it is the case for the logical connectives, but rather from
a certain form of sequence of equations which Statman calls
{\em computation\/}, and whose normal form is called {\em direct computation\/}.
With the formulation of a proof theory for the `logical connective' of
propositional equality we wish to analyse equality reasoning into its
basic components: rewrites, on the one hand, and statements about the existence
of rewrites, on the other hand. This type of analysis came to the surface
in the context of constructive type theory and the Curry--Howard functional
interpretation.

\paragraph*{Martin-L\"of's Identity type.}
There has been essentially two approaches to the problem of characterizing a
proof theory for propositional equality, both of which originate in
P.\ Martin-L\"of's work on {\em Intuitionistic Type Theory\/}: the intensional
\cite{Martin-Lof75:itt} and the extensional \cite{Martin-Lof82:itt,martinlof84}
formulations.

\paragraph*{The extensional version.}
In his \cite{Martin-Lof82:itt} and \cite{martinlof84}
presentations of {\em Intuitionistic Type Theory\/} P.\ Martin-L\"of defines the
type of {\em extensional\/} propositional equality `${\tt Id}$' (here called
`${\tt Id}^{ext}$') as:

\bigskip

\noindent ${\tt Id}^{ext}$-{\it formation\/}
$$\displaystyle{{A\ type \qquad a:A \qquad b:A} \over
{{\tt Id}_A^{ext}(a,b)\ type}}$$

\noindent ${\tt Id}^{ext}$-{\it introduction\/}
$$\displaystyle{{a=b:A} \over {{\tt r}:{\tt Id}_A^{ext}(a,b)}}$$

\noindent ${\tt Id}^{ext}$-{\it elimination\/}\footnote{The set of rules given
in \cite{Martin-Lof82:itt} contained the additional {\it elimination\/} rule:
$$\displaystyle{{c:{\tt Id}_A^{ext}(a,b) \qquad d:C({\tt r}/z)} \over
{{\tt J}(c,d):C(c/z)}}$$
which may be seen as reminiscent of the previous {\em intensional\/} account of
propositional equality \cite{Martin-Lof75:itt}.}
$$\displaystyle{{c:{\tt Id}_A^{ext}(a,b)} \over {a=b:A}}$$

\noindent ${\tt Id}^{ext}$-{\it equality\/}
$$\displaystyle{{c:{\tt Id}_A^{ext}(a,b)} \over {c={\tt r}:{\tt Id}_A^{ext}(a,b)}}$$

Note that the above account of propositional equality does not `keep track of
all proof steps': both in the ${\tt Id}^{ext}$-{\it introduction\/} and in the
${\tt Id}^{ext}$-{\it elimination\/} rules there is a considerable loss of
information concerning the deduction steps. While in the
${\tt Id}^{ext}$-{\it introduction\/} rule the `$a$' and the `$b$' do not appear
in the `trace' (the label/term alongside the logical formula/type), the latter
containing only the canonical element `${\tt r}$', in the rule of
${\tt Id}^{ext}$-{\it elimination\/} all the trace that might be recorded in the
term `$c$' simply disappears from label of the conclusion. If by
`intensionality' we understand a feature of a logical system which identifies as
paramount the concern with issues of {\em context\/} and {\em provability\/},
then it is quite clear that any logical system containing
${\tt Id}^{ext}$-type can hardly be said to be `intensional': as we have said
above, neither its {\it introduction\/} rule nor its {\it elimination\/} rule
carry the necessary {\em contextual\/} information from the premise to the
conclusion.


\paragraph*{The intensional version.}
Another version of the propositional equality, which has its origins in
Martin-L\"of's early accounts of {\em Intuitionistic Type Theory\/}
\cite{Martin-Lof75:models,Martin-Lof75:itt}, and is apparently in the most
recent, as yet unpublished, versions of type theory, is defined in
\cite{TroelstravanDalen88} and \cite{Nordstrometal90}. In a section dedicated to
the {\em intensional vs.\ extensional\/} debate, \cite{TroelstravanDalen88} (p.633)
says that:
\begin{quote}
``Martin-L\"of has returned to an intensional point of view, as in Martin-L\"of
(1975), that is to say, $t=t' \in A$ is understood as ``$t$ and $t'$ are
definitionally equal''. As a consequence the rules for identity types have to be
adapted.''
\end{quote}
If we try to combine the existing accounts of the {\em intensional\/} equality
type `${\tt Id}_A$' \cite{Martin-Lof75:itt,TroelstravanDalen88,Nordstrometal90},
here denoted `${\tt Id}^{int}$', the rules will look like:

\bigskip

\noindent ${\tt Id}^{int}$-{\it formation\/}
$$\displaystyle{{A\ type \qquad a: A \qquad b:A} \over
{{\tt Id}_A^{int}(a,b)\ type}}$$

\noindent ${\tt Id}^{int}$-{\it introduction\/}
$$\displaystyle{{a:A} \over {{\tt r}(a):{\tt Id}_A^{int}(a,a)}} \qquad
\displaystyle{{a=b:A} \over {{\tt r}(a):{\tt Id}_A^{int}(a,b)}}$$

\noindent ${\tt Id}^{int}$-{\it elimination\/}
$$\textstyle{{\displaystyle{\ \atop {a:A\quad b:A\quad c:{\tt Id}_A^{int}(a,b)}} \quad
\textstyle{{[x:A]} \atop {d(x):C(x,x,{\tt r}(x))}} \quad
\textstyle{{[x:A,y:A,z:{\tt Id}_A^{int}(x,y)]} \atop {C(x,y,z)\ type}}} \over
{{\tt J}(c,d):C(a,b,c)}}$$

\noindent ${\tt Id}^{int}$-{\it equality\/}
$$\displaystyle{{\displaystyle{\ \atop {a:A}}\ 
\displaystyle{{[x:A]} \atop {d(x):C(x,x,{\tt r}(x))}} \ 
\displaystyle{{[x:A,y:A,z:{\tt Id}_A^{int}(x,y)]} \atop {C(x,y,z)\ type}}} \over
{{\tt J}({\tt r}(a),d(x))=d(a/x):C(a,a,{\tt r}(a))}}$$

With slight differences in notation, the `adapted' rules for identity type given
in \cite{TroelstravanDalen88} and \cite{Nordstrometal90} resembles the one given
in \cite{Martin-Lof75:itt}. It is called {\em intensional\/} equality because
there remains no direct connection between judgements like `$a=b:A$' and
`$c:{\tt Id}_A^{int}(a,b)$'.

\paragraph*{A labelled proof theory for propositional equality.}
Now, it seems that an alternative formulation of propositional equality within
the functional interpretation, which will be a little more elaborate than the
extensional ${\tt Id}_A^{ext}$-type, and simpler than the intensional
${\tt Id}_A^{int}$-type, could prove more convenient from the point of view of the
`logical interpretation'. It seems that whereas in the former we have a
considerable loss of information in the ${\tt Id}^{ext}$-{\it elimination\/}, in the sense that propositional equality and definitional equality are collapsed into one, in
the latter we have an ${\tt Id}^{int}$-{\it elimination\/} too heavily loaded
with (perhaps unnecessary) information.  If, on the one hand, there is an
{\em over\/}explicitation of information in ${\tt Id}^{int}$, on the other
hand, in ${\tt Id}^{ext}$ we have a case of {\em under\/}explicitation.  With
the formulation of a proof theory for equality via labelled natural deduction
we wish to find a middle ground solution between those two extremes.

\section{Labelled deduction}

The functional interpretation of logical connectives via deductive systems
which use some sort of labelling mechanism
\cite{martinlof84,ldslivro,extending} can be seen as the basis for
a general framework characterizing logics via a clear separation between a
functional calculus on the {\em labels\/}, i.e.\ the referents (names
of individuals, expressions denoting the record of proof steps used to arrive at
a certain formula, names of
`worlds', etc.) and a logical calculus on the formulas. The key idea is to make
these two dimensions as harmonious as possible, i.e.\ that the functional calculus on
the labels matches the logical calculus on the formulas at least in the sense
that to every abstraction on
the variables of the functional calculus there corresponds a discharge of an
assumption-formula of the logical calculus. One aspect of such interpretation
which stirred much discussion in the literature of the past ten years or so,
especially in connection with {\em Intuitionistic Type Theory\/}
\cite{martinlof84}, was that of whether the logical connective of
propositional equality ought to be dealt with `extensionally' or
`intensionally'. Here we attempt to formulate what appears to be a middle ground
solution, in the sense that the intensional aspect is dealt with in the
functional calculus on the labels, whereas the extensionality is kept to the
logical calculus. We also intend to demonstrate that the connective of
propositional equality (cf.\ Aczel's \cite{Aczel80} `${\tt Id}$') needs to be dealt with
in a similar manner to `Skolem-type' connectives (such as disjunction and
existential quantification), where notions like {\em hiding\/}, {\em choice\/}
and {\em dependent variables\/} play crucial r\^oles.

\subsection{Identifiers for (compositions of) equalities}
In the functional interpretation, where a functional calculus on the labels go
hand in hand with a logical calculus on the formulas, we have a classification
of equalities, whose identifications are carried along as part of the deduction:
either $\beta$-, $\eta$-, $\xi$-, $\mu$- or $\alpha$- equality will have been part of
an expression labelling a formula containing `${\tt Id}$'. There one finds the key
to the idea of `hiding' in the {\it introduction\/} rule, and opening local
(Skolem-type) assumptions in the {\it elimination\/} rule. (Recall that in the
case of disjunction we also have alternatives: either into the left disjunct, or
into the right disjunct.) So, we believe that it is not unreasonable to start
off the formalization of propositional equality with the parallel to the
disjunction and existential cases in mind. Only, the witness of the type of
propositional equality are not the `$a$'s and `$b$'s of `$a=b:A$', but the
actual (sequence of) equalities ($\beta$-, $\eta$-, $\xi$-, $\alpha$-) that might
have been used to arrive at the judgement `$a =_s b:A$' (meaning `$a=b$'
{\em because\/} of `$s$'), `$s$' being a sequence made up of $\beta$-, $\eta$-,
$\xi$- and/or $\alpha$-equalities, perhaps with some of the general equality
rules of reflexivity, symmetry and transitivity. So, in the {\it introduction\/}
rule of the type we need to form the canonical proof as if we were
{\em hiding\/} the actual sequence. Also, in the rule of {\it elimination\/} we
need to open a new local assumption introducing a new variable denoting a
possible sequence as a (Skolem-type) new constant. That is, in order to
eliminate the connective `${\tt Id}_A$' (i.e.\ to deduce something from a proposition 
like `${\tt Id}_A(a,b)$'), we start by choosing a new variable to
denote the reason why the two terms are equal: `let $t$ be an expression
(sequence of equalities) justifying the equality between the terms'. If we then
arrive at an arbitrary formula `$C$' labelled with an expression where the
$t$ still occurs free, then we can conclude that the same $C$ can be
obtained from the ${\tt Id}$-formula regardless of the identity of the chosen $t$,
meaning that the label alongside $C$ in the conclusion will have been
abstracted from the free occurrences of $t$.

Observe that now we are still able to `keep track' of all proof steps (which
does not happen with Martin-L\"of's
${\tt Id}_A^{ext}$-type) \cite{Martin-Lof82:itt,martinlof84}, and we have
an easier formulation (as compared with Martin-L\"of's
${\tt Id}_A^{int}$-type) \cite{Martin-Lof75:itt} of how to perform the
{\it elimination\/} step.

\subsection{The proof rules}
In formulating the propositional equality connective, which we shall identify by
`${\tt Id}$', we shall keep the pattern of inference rules essentially the same as
the one used for the other logical connectives (as in, e.g.\ \cite{ldsexistential}), and we shall provide an
alternative presentation of propositional equality as follows:

\bigskip

\noindent ${\tt Id}$-{\it formation\/}
$$\displaystyle{{A\ type \qquad a:A \qquad b:A} \over
{{\tt Id}_A(a,b)\ type}}$$

\medskip

\noindent ${\tt Id}$-{\it introduction\/}
$$\displaystyle{{a=_s b:A} \over {s(a,b):{\tt Id}_A(a,b)}}\qquad \displaystyle{{a=_s b:A\quad a=_t b:A\quad s=_z t:{\tt Id}_A(a,b)} \over {s(a,b)=_{\xi(z)} t(a,b):{\tt Id}_A(a,b)}}$$
(Notice that the $\xi$ rule for ${\tt Id}_A$ has an extra hypothesis, which, though apparently circular, is concerned with making sure that not all sequences of rewrites from $a$ to $b$ are definitionally equal: in order to be declared $\xi$-equal, two sequences need to be equal from some other reason.)

\medskip 

\noindent ${\tt Id}$-{\it elimination\/}
$$\displaystyle{{\displaystyle{\ \atop {p:{\tt Id}_A(a,b)}} \qquad
\displaystyle{{[a=_t b:A]} \atop {d(t):C}}} \over
{{\tt J}(p,\acute{t}d(t)):C}}
\qquad
\displaystyle{{\displaystyle{\ \atop {p=_rq:{\tt Id}_A(a,b)}} \qquad
\displaystyle{{[a=_t b:A]} \atop {d(t):C}}} \over
{{\tt J}(p,\acute{t}d(t))=_{\mu(r)}{\tt J}(q,\acute{t}d(t)):C}}$$

\noindent ${\tt Id}$-{\it reduction\/}
$$\displaystyle{{\displaystyle{{a=_s b:A} \over {s(a,b):{\tt Id}_A(a,b)}}{\tt Id}\mbox{\it -intr\/} \qquad
\displaystyle{{[a=_t b:A]} \atop {d(t):C}}} \over
{{\tt J}(s(a,b),\acute{t}d(t)):C}}{\tt Id}\mbox{\it -elim\/} \qquad
\triangleright_\beta \qquad
\displaystyle{{a=_s b:A} \atop {d(s/t):C}}$$
giving rise to the equality
$${\tt J}(s(a,b),\acute{t}d(t))=_\beta d(s/t):C$$

\noindent ${\tt Id}$-{\it induction\/}
$$\displaystyle{{\displaystyle{\ \atop {e:{\tt Id}_A(a,b)}} \qquad
\displaystyle{{[a=_t b:A]} \over {t(a,b):{\tt Id}_A(a,b)}}{\tt Id}\mbox{\it -intr\/}} \over
{{\tt J}(e,\acute{t}t(a,b)):{\tt Id}_A(a,b)}}{\tt Id}\mbox{\it -elim\/} \qquad \triangleright_\eta \qquad
 e:{\tt Id}_A(a,b)$$
giving rise to the equality
$${\tt J}(e,\acute{t}t(a,b))=_\eta e:{\tt Id}_A(a,b)$$
where `$\acute{\ }$' is an abstractor which binds the occurrences of the (new)
variable `$t$' introduced with the local assumption `$[a=_t b:A]$'
as a kind of `Skolem'-type constant denoting the (presumed) `reason' why `$a$'
was assumed to be equal to `$b$'. (Recall the Skolem-type procedures of
introducing new local assumptions in order to allow for the elimination of
logical connectives where the notion of `hiding' is crucial, e.g.\ disjunction
and existential quantifier -- in \cite{ldsexistential}.)

Now, having been defined as a `Skolem'-type connective, `${\tt Id}$' needs to have
a conversion stating the non-interference of the newly opened branch (the local
assumption in the ${\tt Id}$-{\it elimination\/} rule) with the main branch. Thus,
we have:

\bigskip

\noindent ${\tt Id}$-({\it permutative\/}) {\it reduction\/}
$$\displaystyle{{\displaystyle{{\displaystyle{\ \atop {e:{\tt Id}_A(a,b)}} \quad
\displaystyle{{[a =_t b:A]} \atop {d(t):C}}} \over
{{\tt J}(e,\acute{t}d(t)):C}}} \over
{w({\tt J}(e,\acute{t}d(t))):W}}{\sf r}
\quad \triangleright_\zeta \quad
\displaystyle{{\displaystyle{\ \atop {e:{\tt Id}_A(a,b)}} \quad
\displaystyle{{\displaystyle{{[a =_t b:A]} \atop {d(t):C}}} \over
{w(d(t)):W}}{\sf r}} \over
{{\tt J}(e,\acute{t}w(d(t))):W}}$$
provided $w$ does not disturb the existing dependencies in the term $e$ (the
main branch), i.e.\ provided that rule `{\sf r}' does not discharge any
assumption on which `${\tt Id}_A(a,b)$' depends. The corresponding
$\zeta$-equality is:
$$w({\tt J}(e,\acute{t}d(t)))=_\zeta {\tt J}(e,\acute{t}w(d(t)))$$
The equality indicates that the operation $w$ can be pushed inside the
$\acute{\ }$-abstraction term, provided that it does not affect the dependencies of
the term $e$.

Since we are defining the logical connective `${\tt Id}$' as a connective which
deals with singular terms, where the `witness' is supposed to be hidden, we
shall not be using direct {\it elimination\/} like Martin-L\"of's
${\tt Id}^{ext}$-{\it elimination\/}. Instead, we shall be using the following
${\tt Id}$-{\it elimination\/}:
$$\displaystyle{{\displaystyle{\ \atop {e:{\tt Id}_A(a,b)}} \qquad
\displaystyle{{[a=_t b:A]} \atop {d(t):C}}} \over
{{\tt J}(e,\acute{t}d(t)):C}}\qquad
\displaystyle{{\displaystyle{\ \atop {e=_s f:{\tt Id}_A(a,b)}} \qquad
\displaystyle{{[a=_t b:A]} \atop {d(t):C}}} \over
{{\tt J}(e,\acute{t}d(t))=_{\mu\cdot s}{\tt J}(f,\acute{t}d(t)):C}}$$
The {\it elimination\/} rule involves the introduction of a new local assumption
(and corresponding variable in the functional calculus), namely `$[a=_t b:A]$'
(where `$t$' is the new variable) which is only discharged (and `$t$' bound) in
the conclusion of the rule. The intuitive explanation would be given in the
following lines. In order to eliminate the equality ${\tt Id}$-connective, where
one does not have access to the `reason' (i.e.\ a sequence of `$\beta$',
`$\eta$', `$\xi$' or `$\zeta$' equalities) why the equality holds because
`${\tt Id}$' is supposed to be a connective dealing with singular terms (as are
`$\lor$' and `$\exists$'), in the first step one has to open a new local
assumption supposing the equality holds because of, say `$t$' (a new variable).
The new assumption then stands for `let $t$ be the unknown equality'. If a third
(arbitrary) statement can be obtained from this new local assumption via an
unspecified number of steps which does not involve any binding of the new
variable `$t$', then one discharges the newly introduced local assumption
binding the free occurrences of the new variable in the label alongside the
statement obtained, and concludes that that statement is to be labelled by the
term `${\tt J}(e,\acute{t}d(t))$' where the new variable (i.e.\ $t$) is bound
by the `$\acute{\ }$'-abstractor.

Another feature of the ${\tt Id}$-connective which is worth noticing at this stage
is the equality under `$\xi$' of all its elements (see second
{\it introduction\/} rule). This does not mean that the labels serving as
evidences for the ${\tt Id}$-statement are all identical to a constant
(cf.\ constant `{\tt r}' in Martin-L\"of's ${\tt Id}_{ext}$-type), but simply
that if two (sequences of) equality are obtained as witnesses of the equality
between, say `$a$' and `$b$' of domain $A$, then they are taken to be equal
under $\xi$-equality. It would not seem unreasonable to think of the
${\tt Id}$-connective of propositional equality as expressing the proposition
which, whenever true, indicates that the two elements of the domain concerned
are equal under some (unspecified, {\em hidden\/}) composition of definitional
equalities. It is as if the proposition points to the existence of a term
(witness) which depends on both elements and on the kind of equality judgements
used to arrive at its proof. So, in the logical side, one forgets about what was
the actual witness. Cf.\ the existential generalization:
$$\displaystyle{{F(t)} \over {\exists x.F(x)}}$$
where the actual witness is in fact `abandoned'. Obviously, as we are interested
in keeping track of relevant information introduced by each proof step, in
labelled natural deduction system the witness is not abandoned, but is carried
over as an unbounded name in the label of the corresponding conclusion formula.
$$\displaystyle{{t:A \qquad f(t):F(t)} \over {\varepsilon x.(f(x),t):\exists x^ A.F(x)}}$$
Note, however, that it is carried along {\em only\/} in the functional side, the
logical side not keeping any trace of it at all.

Now, notice that if the functional calculus on the labels is to match the
logical calculus on the formulas, than we must have the resulting label on the
left of the `$\triangleright_\beta$' as $\beta$-convertible to the
concluding label on the right. So, we must have the convertibility equality:
$${\tt J}(s(a,b),\acute{t}d(t))=_\beta d(s/t):C$$
The same holds for the $\eta$-equality:
$${\tt J}(e,\acute{t}t(a,b))=_\eta e:{\tt Id}_A(a,b)$$

Parallel to the case of disjunction, where two different {\tt constructors}
distinguish the two alternatives, namely `$i$' and `$j$', we
here have any (sequence of) equality identifiers (`$\beta$', `$\eta$', `$\mu$',
`$\xi$', etc.) as {\tt constructor}s of proofs for the ${\tt Id}$-connective.
They are meant to denote the alternatives available.

\paragraph*{General rules of equality.}
Apart from the already mentioned `constants' (identifiers) which compose the reasons for
equality (i.e.\ the indexes to the equality on the functional calculus), it is
reasonable to expect that the following rules are taken for granted:
{\it reflexivity\/}, {\it symmetry\/} and {\it transitivity\/}.

\paragraph*{Substitution without involving quantifiers.}
We know from logic programming, i.e.\ from the theory of unification, that
substitution can take place even when no quantifier is involved. This is
justified when, for some reason a certain referent can replace another under
some condition for identifying the one with the other.

Now, what would be counterpart to such a `quantifier-less' notion of
substitution in a labelled natural deduction system. Without the appropriate
means of handling equality (definitional and propositional) we would hardly be
capable of finding such a counterpart. Having said all that, let us think of
what we ought to do at a certain stage in a proof (deduction) where the
following two premises would be at hand:
$$a=_g y:A \qquad \mbox{ and } \qquad f(a):P(a)$$
We have that $a$ and $y$ are equal (`identifiable') under some arbitrary
sequence of equalities (rewrites) which we name $g$. We also have that the
predicate formula $P(a)$ is labelled by a certain functional expression
$f$ which depends on $a$. Clearly, if $a$ and $y$ are `identifiable', we would
like to infer that $P$, being true of $a$, will also be true of $y$. So,
we shall be happy in inferring (on the logical calculus) the formula
$P(y)$. Now, given that we ought to compose the label of the conclusion
out of a composition of the labels of the premises, what label should we insert
alongside $P(y)$? Perhaps various good answers could be given here, but
we shall choose one which is in line with our `keeping record of what (relevant)
data was used in a deduction'. We have already stated how much importance we
attach to names of individuals, names of formula instances, and of course, what
kind of deduction was performed (i.e.\ what kind of connective was introduced
or eliminated). In this section we have also insisted on the importance of, not
only `classifying' the equalities, but also having variables for the kinds of
equalities that may be used in a deduction. Let us then formulate our rule of 
`quantifier-less' substitution as:
$$\displaystyle{{a=_g y:A \qquad f(a):P(a)} \over {g(a,y)\cdot f(a):P(y)}}$$
which could be explained in words as follows: if $a$ and $y$ are `identifiable'
due to a certain $g$, and $f(a)$ is the evidence for $P(a)$, then let the
composition of $g(a,y)$ (the label for the propositional equality between $a$
and $y$) with $f(a)$ (the evidence for $P(a)$) be the evidence for
$P(y)$.

By having this extra rule of substitution added to the system of rules of
inference, we are able to validate one half of the so-called `Leibniz's law',
namely:
$$\forall x^A\forall y^A.({\tt Id}_A(x,y)\to(P(x)\to P(y)))$$

\paragraph*{The {\em LND} equational fragment.}
As we already mentioned, in the {\em LND\/} equational logic, the equations 
have an index (the {\em reason}) which keeps all proof steps. The {\em reasons}
 is defined by the kind of rule used in the proof and the equational axioms ({\em definitional equalities\/}) of the system. 
The rules are divided into the
 following classes: (i) general rules; (ii) subterm  substitution rule;
(iii) $\xi$- and $\mu$-rules.

Since the {\em LND\/} system is based on the Curry--Howard isomorphism
\cite{howard80}, terms represent proof constructions, thus proof transformations
correspond to equalities between terms. In this way, the {\em LND} equational
logic can deal with equalities between {\em LND\/} proofs. The proofs in the
{\em LND\/} equational fragment which deals with equalities between deductions
are built from the basic proof transformations for the {\em LND\/} system, given
in \cite{ldsintro,ldsexistential,lndlivro}.
These basic proof transformations form an equational system, composed by
{\em definitional equalities\/} ($\beta$, $\eta$ and $\zeta$).

\paragraph*{General rules.}

\begin{definition}[equation]
An equation in $LND_{EQ}$ is of the form:
$$s=_r t:A$$
where $s$ and $t$ are terms, $r$ is the identifier for the {\em rewrite reason\/},
and $A$ is the type (formula).
\end{definition}

\begin{definition}[system of equations]
A system of equations $S$ is a set of equations:
$$\{s_1=_{r_1}t_1:A_1, \ldots, s_n=_{r_n}t_n:A_n\}$$
where $r_i$ is the {\em rewrite reason\/} identifier for the $i$th equation in $S$.
\end{definition}

\begin{definition}[rewrite reason]
Given a system of equations $S$ and an equation $s=_r t:A$, if $S\vdash s=_r t:A$, i.e.\ there is a deduction/computation of the equation starting
from the equations in $S$, then the {\em rewrite reason\/} $r$ is built up from:
\begin{enumerate}
\item[(i)] the constants for {\em rewrite reasons\/}: $\{\ \rho, \beta, \eta, \zeta\ \}$; 
\item[(ii)] the $r_i$'s;
\end{enumerate}
using the substitution operations:
\begin{enumerate}
\item[(iii)] ${\tt sub}_{\tt L}$;
\item[(iv)] ${\tt sub}_{\tt R}$;
\end{enumerate}
and the operations for building new {\em rewrite reasons\/}:
\begin{enumerate}
\item[(v)] $\sigma$, $\tau$, $\xi$, $\mu$.
\end{enumerate}
\end{definition}

\begin{definition}[general rules of equality]
The general rules for equality (reflexivity, symmetry and transitivity) are
 defined as follows:

$$\begin{array}{lll}
\mbox{\it reflexivity\/} & \mbox{\it symmetry\/} & \mbox{\it transitivity\/} \\
\  & \  & \ \\
\displaystyle{{x:A} \over {x=_\rho x:A}}\quad &
\displaystyle{{x=_t y:A} \over {y=_{\sigma(t)} x:A}}\quad &
\displaystyle{{x=_t y:A \qquad y=_u z:A} \over {x=_{\tau(t,u)} z:A}}
\end{array}$$
\end{definition}

\paragraph*{The ``subterm substitution'' rule.}
Equational logic as usually presented has the following inference rule of 
substitution: 

$$ \displaystyle{ s=t \over  s\theta = t\theta}$$
where $\theta$ is a substitution.

Note that the substitution $\theta$ ``appeared'' in the conclusion of the
rule.  As rightly pointed out by Le~Chenadec in \cite{chenadec89}, from the
viewpoint of the subformula property (objects in the conclusion of some
inference should be subobjects of the premises), this rule is unsatisfactory.
He then defines two rules:

$${IL}\displaystyle{M=N \quad C[N]=O \over C[M] = O} \qquad {IR}\displaystyle{M=
C[N] \quad N=O \over M=C[O]}$$
where  $M$, $N$ and $O$ are terms and the context $C[\_]$ is adopted in order 
to distinguish subterms.

In \cite{deOliveira2} we have formulated an inference rule 
called ``subterm substitution'' which deals in a {\em explicit way\/}\footnote{ In \cite{ldsintro} de Queiroz and Gabbay recall Girard, who describes the
intimate connections between constructivity and explicitation, and claim that
``...one of the aims of inserting a label alongside formulas (accounting for 
the steps made to arrive at each particular point in the deduction) is exactly
 that of making {\em explicit} the use of formulas (and instances of formulas 
and individuals) throughout a deduction ...''} with substitutions.
In fact, the {\em LND\/} \cite{ldsintro,lndlivro} can be seen as an enriched system
which brings to the object language terms, and now substitutions.

\begin{definition}[subterm substitution]
The rule of ``subterm substitution'' is framed as follows:

$$\displaystyle{x =_r {\cal C}[y]: A \quad y =_s u : A' \over
x =_{{\tt sub}_{\tt L}(r,s)} {\cal C}[u]: A}
\qquad
\displaystyle{x =_r w : A' \quad {\cal C}[w]=_s u : A \over
{\cal C}[x]=_{{\tt sub}_{\tt R}(r,s)} u : A}$$
where ${\cal C}$ is the context in which the subterm detached by `$[\; ]$' 
appears and $A'$ could be a subdomain of $A$, equal to $A$ or
 disjoint to $A$. (${\cal C}[u]$ is the result of replacing all occurrences of $y$ by $u$ in ${\tt C}$.\footnote{We should like to thank an anonymous referee who pointed out the ambiguity which would remain in case this condition is not made clear.} 
\end{definition}

The symbols ${\tt sub}_{\tt L}$ and ${\tt sub}_{\tt R}$ denote in which side (L -- {\em left} or R -- {
\em right}) is the premiss that contains the subterm to be substituted.

Note that the transitivity rule previously defined can be seen as a special 
case for this rule when $A' =  A$ and the context ${\cal C}$ is empty.

\paragraph*{The $\xi$- and $\mu$-rules.}
In the Curry--Howard ``formulae-as-types'' interpretation \cite{howard80}, the $\xi$-rule\footnote{The $\xi$-rule is the formal counterpart to Bishop's
 constructive principle of definition of a set \cite{bishop67} (page 2) which
 says: ``To define a set we prescribe, at least implicitly, what 
we have (the constructing intelligence) must to do in order to construct an
 element of the set, and what we must do to show that two elements of the set
 are equal.'' Cf.\ also \cite{bishop67} (page 12) Bishop defines a 
product of set as ``The {\em cartesian product}, or simply 
{\em product\/}, $X \equiv X_1 \times \ldots \times X_n$ of sets 
$X_1,X_2, \ldots , X_n$ is defined to be the set of all ordered n-tuples 
($x_1, \ldots , x_n$) and ($y_1, \ldots , y_n$) of $X$ are equal if the 
coordinates $x_i$ and $y_i$ are equal for each $i$.'' 
 See also \cite{martinlof84} (p.8): ``... a set A is
 defined by prescribing how a canonical element of A is formed as well as how
 two equal canonical elements of A are formed.''
 We also know from the theory of Lambda Calculus the definition of $\xi$-rule,
 see e.g.\ \cite{barend84} (pp.\ 23 and 78): ``$\xi : \; M = N 
\Rightarrow \lambda x . M = \lambda x . N$''} states when two {\em 
canonical\/} elements are equal, and the $\mu$-rule\footnote{The $\mu$-rule is 
also defined in the theory of Lambda Calculus, see e.g.\ \cite{mitchell}: 
``The equational axioms and inference rules are as follows, 
where $[N/x ]M$ denotes substitution of $N$ for $x$ in $M$. $\ldots$  
 $$(\mu) \qquad \displaystyle{ \Gamma \vartriangleright M_1 = M_2 : \sigma \Rightarrow \tau 
 \quad \Gamma \vartriangleright N_1 =N_2 : \sigma \over \Gamma \vartriangleright M_1 N_1 = M_2 N_2 : \tau}\mbox{''}$$
and is divided into two equalities $\mu$ and $\nu$ in \cite{hindleyseldin} (p.66):
$$(\mu)\displaystyle{{M=M'}\over{NM=NM'}}\qquad\qquad
(\nu)\displaystyle{{M=M'}\over{MN=M'N}}$$}
states when two {\em noncanonical\/} elements are equal. 
So, each introduction rule for the {\em LND\/} system has associated to it a 
$\xi$-rule and each elimination rule has a related $\mu$-rule. For instance, 
the $\xi$-rule and $\mu$-rule for the connective $\land$ are defined as follows:
$$\begin{array}{ll} 
\displaystyle{
{x=_u y:A \qquad s:B}
}\over
\displaystyle{
{\langle x,s \rangle=_{\xi_1 (u)} \langle y,s \rangle:A \wedge B}
}\qquad&
\displaystyle{
{x:A \qquad s=_v t:B}
}\over
\displaystyle{
{\langle x,s \rangle=_{\xi_2 (v)} \langle x,t \rangle:A \wedge B}
}\end{array}$$
$$\begin{array}{ll} 
\displaystyle{{x=_r y:A \wedge B} \over {{FST}(x) =_{\mu_1 (r)}{
FST}(y) : A}}\qquad\ & 
\displaystyle{{x=_ry:A \wedge B} \over {{SND}(x) =_{\mu_2 ( r)}{SND}
(y): B}}
\end{array}$$
In the Appendix we give a list, for each type-forming operator, of rules of definition, in the style of Bishop, Curry-Howard, and Martin-L\"of, of what elements the type contains, and when two elements are defined to be equal.

\paragraph*{Term rewriting system for {\em LND\/} with equality.}
In \cite{deOliveira3} we have proved termination and confluence for the
rewriting system arising out of the proof rules given for the proposed natural
deduction system for equality.

The idea is to analyse all possible occurrences of redundancies in proofs which involve the
rules of rewriting, and the most obvious case is the nested application of the rule of symmetry.
But there are a number of cases when the application of rewriting rules is redundant, but
which is not immediately obvious that there is a redundancy. Take, for instance, the following case:
\begin{definition}[reductions involving $\tau$]
$$\displaystyle{ x =_r y : {A} \quad y =_{\sigma(r)} x :
 {A} \over x =_{\tau(r,\sigma(r))} x : {A}} \quad 
\triangleright_{tr}  \quad x =_{\rho} x : {A}$$
$$\displaystyle{ y =_{\sigma (r)} x : {A} \quad x=_r y 
:  {A} \over y =_{\tau(\sigma(r),r)} y : {A}} \quad \triangleright_{tsr} \quad  y=_{\rho} y : {A}$$
$$\displaystyle{ u=_r v : {A} \quad v=_{\rho} v :  
{A} \over u =_{\tau(r,\rho)} v : {A}} \quad 
\triangleright_{trr}  \quad  u =_r v : {A}$$
$$\displaystyle{ u=_{\rho} u : {A} \quad u=_r v : 
{A} \over u=_{\tau(\rho, r)} v : {A}} \quad 
\triangleright_{tlr}  \quad u=_r v : {A}$$
Associated rewriting rule over the reason:\\
$\tau(r,\sigma(r))\triangleright_{tr}\rho$\\
$\tau(\sigma(r),r) \triangleright_{tsr}\rho$\\
$\tau(r,\rho) \triangleright_{trr} r$\\
$\tau(\rho, r) \triangleright_{tlr} r$.\\
\end{definition} 

Below is another less obvious case of ocurrence of redundancy:
\begin{definition}\ \\
$\displaystyle{\displaystyle{\displaystyle{\displaystyle{ \atop } \atop 
}  \atop a: { A} } \quad \displaystyle{\displaystyle{ 
\displaystyle{ {[ x: {A} ]} \atop \vdots} \atop { b(x) =_r g(x) : {B}}}
\over \lambda x.b(x) =_{\xi ( r)} \lambda x . g(x) : { A} \to {B}}\to \mbox{\it -intr} \over
{APP} ( \lambda x. b(x) , a) =_{ \nu ( \xi ( r))} {APP}  
(\lambda x. g(x) , a) : {B}}\to\mbox{\it -elim}$

\smallskip

\hfill{$\triangleright_{mxl} \quad \displaystyle{{a:A}\atop{b(a/x) =_{r} g(a/x) : {B}}}$}\\
\ \\
Associated rewriting rule:\\
$ \nu ( \xi ( r)) \triangleright_{mxl} r$.
\end{definition}

\medskip

As an example:
\begin{example}\ \\
$\displaystyle{\displaystyle{\displaystyle{ \atop x=_r y :{A}} \over
 {i} (x) =_{\xi_1 (r)} i(y) : {A} + {B}} \quad 
\displaystyle{\displaystyle{ x =_r y : {A} \over y=_{\sigma (r)} x :
 {A}} \over {i} (y) =_{\xi_1(\sigma (r))} {i} (x) : {A}
 + {B}} \over {i} (x) =_{\tau ( \xi_1 (r), \xi_1 (\sigma 
(r)))} {i} (x) : {A} + {B}} $

\hfill{$\triangleright_{tr} \quad \displaystyle{{x=_r y:A}\over{x=_{\xi_1(r)}y:A+B}}$}
\ \\
Associated rewriting:\\
$\tau ( \xi_1 (r), \xi_1 (\sigma (r))) \triangleright_{tr} \xi_1 (r)$.
\end{example}


\begin{definition}[reductions involving $\rho$ and $\sigma$]
$$ \displaystyle{ x=_\rho x : {A} \over x =_{\sigma(\rho)} x : {A}}  \quad \triangleright_{sr} \quad  x =_\rho x : {A}$$
$$ \displaystyle{\displaystyle{ x=_r y : {A} \over y =_{\sigma(r)
} x : {A}} \over x =_{\sigma(\sigma(r)) }y:A}  \quad \triangleright_{sr} \quad
 x=_r y : {A}$$
Associated rewritings:\\
$\sigma(\rho)\triangleright_{sr} \rho$ \\
$\sigma(\sigma(r))\triangleright_{sr} r$ \\
\end{definition}


\begin{definition}[reductions involving $\tau$]
$$\displaystyle{ x =_r y : {A} \quad y =_{\sigma(r)} x :
 {A} \over x =_{\tau(r,\sigma(r))} x : {A}} \quad
\triangleright_{tr}  \quad x =_{\rho} x : {A}$$
$$\displaystyle{ y =_{\sigma (r)} x : {A} \quad x=_r y
:  {A} \over y =_{\tau(\sigma(r),r)} y : {A}} \quad \triangleright_{tsr} \quad  y=_{\rho} y : {A}$$
$$\displaystyle{ u=_r v : {A} \quad v=_{\rho} v :
{A} \over u =_{\tau(r,\rho)} v : {A}} \quad
\triangleright_{trr}  \quad  u =_r v : {A}$$
$$\displaystyle{ u=_{\rho} u : {A} \quad u=_r v :
{A} \over u=_{\tau(\rho, r)} v : {A}} \quad
\triangleright_{tlr}  \quad u=_r v : {A}$$
Associated rewritings:\\
$\tau(r,\sigma(r))\triangleright_{tr} \rho$ \\
$\tau(\sigma(r),r)\triangleright_{tsr} \rho$ \\
$\tau(r,\rho)\triangleright_{tsr} r$ \\
$\tau(\rho,r)\triangleright_{tlr} r$ \\
\end{definition}

Note that the first two reductions identify  the case in which a {\em 
reason\/}  which is part of a rewrite sequence meets its inverse.

These reductions can be generalized to transformations where the reasons $r$
 and $\sigma(r)$ (transf.\ 1 and 2) and $r$ and $\rho$ (transf.\ 3 and 4) appear
 in some context, as illustrated by the following example:

\begin{example}\ \\
$\displaystyle{\displaystyle{\displaystyle{ \atop x=_r y :{A}} \over
 {i} (r) =_{\xi_1 (r)} {i} (y) : {A} + {B}} \quad
\displaystyle{\displaystyle{ x =_r y : {A} \over y=_{\sigma (r)} x :
 {A}} \over {i} (y) =_{\xi_1 (\sigma (r))} {i} (x) : {A}
 + {B}} \over {i} (x) =_{\tau ( \xi_1 (r), \xi_1 (\sigma
(r)))} {i} (x) : {A} + {B}} $

\hfill{$\triangleright_{tr} \quad \displaystyle{\displaystyle{ x : {A} \over x=_\rho x
 : {A}} \over {i}(x) =_{\xi_1 (\rho)} {i} (x) : {A}
 + {B}} $}\\
 Associated rewriting:\\
$\tau ( \xi_1 (r), \xi_1 (\sigma
(r)))\triangleright_{tr} \xi_1(\rho)$ \\
\end{example}
For the general context ${\cal C}[\ ]$:\\
\noindent Associated rewritings:\\
$\tau({\cal C}[r] , {\cal C}[\sigma(r)] ) \triangleright_{tr} {\cal C}[\rho]$\\
$\tau({\cal C}[\sigma(r)] , {\cal C}[r]) \triangleright_{tsr} {\cal C}[\rho]$\\
$\tau({\cal C}[r], {\cal C}[\rho]) \triangleright_{trr} {\cal C}[r]$\\
$\tau({\cal C}[\rho], {\cal C}[r]) \triangleright_{tlr} {\cal C}[r]$

\begin{definition}[substitution rules]
$$\displaystyle{ u =_r {\cal C}[x]:{A} \quad x =_{\rho} x : {A'} \over u =_{{\tt sub_L}(r,\rho)}  {\cal C}[x] : {A}} \quad \triangleright_{slr} \quad u =_r {\cal C}[x]: {A}$$
$$\displaystyle{ x=_{\rho} x : {A'} \quad {\cal C}[x] =_r z : {A} \over {\cal C}[x] =_{{\tt sub_R}(\rho,r)} z :
{A}} \quad \triangleright_{srr} \quad {\cal C}[x] =_r z : {A}$$

$$\displaystyle{\displaystyle{ z =_s {\cal C}[y] : {A}
\quad y =_r w : {A'} \over z =_{{\tt sub_L} (s,r)} {\cal C}[w] :
{D}} \quad \displaystyle{ y =_r w :{A'} \over w =_{\sigma(r)} y :
{D'}} \over z =_{{\tt sub_L} ({\tt sub_L} ( s,r) , \sigma (r))} {\cal C}[y] : {A}} \  \triangleright_{sls} \  z =_s {\cal C}[y] : {A}$$

$$\displaystyle{\displaystyle{  z =_{s} {\cal C}[y] : { A}  \quad   y =_{r} w : {A'} \over z =_{{\tt sub_L} (  s , r)} {\cal C}[w] : {A}} \quad \displaystyle{ y =_r w : {A'} \over w=_{\sigma (r)} y : {A'}} \over z =_{{\tt sub_L} ( {\tt sub_L} ( s, r) ,
\sigma(r))} {\cal C}[y] : {A}} \; \triangleright_{slss} \;  z =_{s} {\cal C}[y]  : {A}$$

$$\displaystyle{ \displaystyle{\displaystyle{ \atop } \atop
x=_s w:  {A'}} \quad  \displaystyle{ \displaystyle{ x=_s w : {A'}
\over  w =_{\sigma (s)} x : {A'}} \quad \displaystyle{ \atop {\cal C}[x] =_{r} z : {A}} \over {\cal C}[w] =_{{\tt sub_R} ( \sigma  (s) , r)} z : {A}} \over  {\cal C}[x] =_{{\tt sub_R}  (s,
{\tt sub_R} ( \sigma (s), r))} z : {A}} \; \triangleright_{srs} \;  {\cal C}[x]  =_{r} z : {A}$$

$$\displaystyle{\displaystyle{ x=_s w : {A'} \over w
=_{\sigma (s)} x : {A'}} \quad \displaystyle{ x=_s w : {A'}
\quad {\cal C}[w] =_r z : {A} \over {\cal C}[x] =_{{\tt sub_R}
(s,r)}  z : {A}} \over {\cal C}[w] =_{{\tt sub_R} ( \sigma (s)
,  {\tt sub_R} (s,r))} z : {A}} \ \triangleright_{srrr} \ {\cal C}[w] =_r
z :  {A}$$

\medskip\noindent
Associated rewritings:\\
${\tt sub_L}({\cal C}[r] , {\cal C}[\rho]) \triangleright_{slr} {\cal C}[r]$\\
${\tt sub_R}({\cal C}[\rho], {\cal C}[r]) \triangleright_{srr} {\cal C}[r]$\\
${\tt sub_L} ( {\tt sub_L} ( s ,{\cal C}[r])  , {\cal C}[\sigma(r)]) \triangleright_{sls} s$\\
${\tt sub_L} ({\tt sub_L} (s, {\cal C}[\sigma(r)]) , {\cal C}[r] ) \triangleright_{slss} s$\\
${\tt sub_R} (s, {\tt sub_R} ({\cal C}[\sigma(s)], r)) \triangleright_{srs} r$\\
${\tt sub_R} ({\cal C}[\sigma(s)], {\tt sub_R} ({\cal C}[s], r)) \triangleright_{srrr} r$\\
\end{definition}


\begin{definition}\ \\
$\beta_{rewr}$-$\times$-{\it reduction}

\noindent $\displaystyle{\displaystyle{x=_r y : { A}  \qquad z : { B}
\over
\langle x,z \rangle =_{\xi_1 (r)} \langle y,z \rangle : {A} \times {B}
}\times \mbox{{\it -intr}}
\over
{FST}( \langle x,z \rangle ) =_{\mu_1 ( \xi_1 ( r))} {FST}(\langle y,z 
\rangle ) : {A} 
}\times \mbox{{\it -elim}}$

\hfill{$\triangleright_{mx2l} \quad x =_r y : {A}$}

\medskip

\noindent $\displaystyle{\displaystyle{x =_rx': { A}  \qquad y=_s z : { B}
\over
\langle x,y \rangle =_{\xi_\land (r,s)} \langle x',z \rangle : {A} \times {B}
}\times \mbox{{\it -intr}}
\over
{FST}( \langle x,y \rangle ) =_{\mu_1 ( \xi_\land( r,s))} {FST}(\langle x',z 
\rangle ) : {A} 
}\times \mbox{{\it -elim}}$

\hfill{$\triangleright_{mx2l} \quad x =_r x' : {A}$}

\medskip

\noindent $\displaystyle{\displaystyle{x=_r y : {A}  \qquad z=_sw  : {B}
\over
\langle x,z \rangle =_{\xi_\land (r,s)} \langle y,w \rangle : {A} \times {B}
}\times \mbox{{\it -intr}}
\over
 {SND} (\langle x, z \rangle ) =_{\mu_2 ( \xi_\land( r,s))} {SND} (\langle
 y,w \rangle ) : {B}
}\times \mbox{\it -elim}$

\hfill{$\triangleright_{mx2r} \quad z =_s w : {B}$}

\medskip

\noindent $\displaystyle{\displaystyle{x: {A}  \qquad z =_s w : {B}
\over
\langle x,z \rangle =_{\xi_2 (s)} \langle x,w \rangle : {A} \times {B}
}\times \mbox{{\it -intr}}
\over
 {SND} (\langle x, z \rangle ) =_{\mu_2 ( \xi_2 ( s))} {SND} (\langle
 x,w \rangle ) : {B}
}\times \mbox{\it -elim}$

\hfill{$\triangleright_{mx2r} \quad z =_s w : {B}$}

\medskip\noindent
Associated rewritings: \\
$\mu_1 ( \xi_1 ( r))\triangleright_{mx2l1} r$\\
$\mu_1 ( \xi_\land (r,s))\triangleright_{mx2l2} r$\\
$\mu_2 ( \xi_\land( r,s))\triangleright_{mx2r1} s$\\
$\mu_2 ( \xi_2 ( s))\triangleright_{mx2r2} s$

\medskip

\noindent $\beta_{rewr}$-$+$-{\it reduction}

\noindent $\displaystyle{{\displaystyle{{a =_r a':{A}} \over 
{i}(a) =_{\xi_1 (r)} {i}(a'):{A} + {B}}+\mbox{
\it -intr\/} \ 
\displaystyle{{[x:{A}]} \atop {f(x) =_s k(x):{C}}} \ 
 \displaystyle{ {[y: {B}]} \atop { g(y) =_u h(y):{C}}}} \over
{{D}({i}(a),\acute{x}f(x), \acute{y}g(y)) =_{\mu
 (\xi_1 (r),s,u)} {D}({i}(a'),\acute{x}k(x), \acute{y}h(y)):{C}}}+\mbox{\it -elim}$

\hfill{$ \triangleright_{mx3l} \quad
\displaystyle{{a =_r a':{A}} \atop {f(a/x) =_s k(a'/x):{C}}}$}

\medskip

\noindent $\displaystyle{{\displaystyle{{b =_r b':{B}} \over {j}(b) =_{\xi_2 (r)} {j} (b'):{A} + {B}}+\mbox{\it -intr\/} \ 
\displaystyle{{[x:{A}]} \atop {f(x) =_s k(x):{C}}} \ 
\displaystyle{ {[y: {B}]} \atop { g(y) =_u h(y):{C}}}} \over
{{D}({j}(b),\acute{x}f(x), \acute{y}g(y)) =_{\mu
 (\xi_2 (r),s,u)} {D}({j}(b'),\acute{x}k(x), \acute{y}h(y)):{C}}}+\mbox{\it -elim}$

\hfill{$\triangleright_{mx3r} \qquad \displaystyle{b =_s b':{B} \atop g(b/y) =_u h(b'/y):{C}}$}

\medskip

\noindent
Associated rewritings:\\
$\mu ( \xi_1 ( r) , s,u)\triangleright_{mx3l} s$\\
$\mu ( \xi_2 ( r), s,u)\triangleright_{mx3r} u$

\medskip








\noindent $\beta_{rewr}$-$\Pi$-{\it reduction}

\noindent $\displaystyle{\displaystyle{\displaystyle{ \atop } \atop a : {A}} 
\quad  \displaystyle{\displaystyle{ [ x : {A} ] \atop f(x) =_r g(x) :  
{B} (x)} \over \lambda x.f(x) = _{\xi (r)} \lambda x.g(x) : 
\Pi  x:{A}.{B} (x)} \over {APP} (\lambda x.f(x), a) 
=_{\nu (\xi (r))} {APP} (\lambda x.g(x) , a) : {B} (a)}$

\hfill{$\triangleright_{mxl} \quad \displaystyle{  a: {A}
  \atop f(a/x) =_r g(a/x) : {B} (a)}$}




\smallskip

\noindent Associated rewriting:\\
$\nu (\xi (r)) \triangleright_{mxl} r$\\

\smallskip

\noindent $\beta_{rewr}$-$\Sigma$-{\it reduction}

\noindent $\displaystyle{\displaystyle{ a=_r a': {A} \quad f(a) : {B}(a) 
\over  \varepsilon x.(f(x),a) =_{\xi_1 (r)} \varepsilon x.(f(x),a') : 
\Sigma  x :{A}.{B} (x)} \quad \displaystyle{ [ t : { A}, 
g(t) :  {B} (t) ] \atop d(g,t) =_s h(g,t) : { C}} \over {E}  
(\varepsilon x.(f(x),a), \acute{g} \acute{t} d(g,t)) = _{\mu (\xi_1 ( r) 
, s) } {E}(\varepsilon x.(f(x),a') , \acute{g} \acute{t}h(g,t)) : 
{ C}}$

\hfill{$\triangleright_{mxr}  
\quad \displaystyle{  a=_r a': {A} \quad f(a) : {B} (a)  
\atop  d(f/g , a/t ) =_s h(f/g , a'/t):{ C}}$}

\smallskip

\noindent $\displaystyle{\displaystyle{ a: {A} \quad f(a)=_r i(a) : {B}(a) 
\over  \varepsilon x.(f(x),a) =_{\xi_2 (r)} \varepsilon x.(i(x),a) : 
\Sigma  x :{A}.{B} (x)} \quad \displaystyle{ [ t : { A}, 
g(t) :  {B} (t) ] \atop d(g,t) =_s h(g,t) : { C}} \over {E}  
(\varepsilon x.(f(x),a), \acute{g} \acute{t} d(g,t)) = _{\mu (\xi_2 ( r) 
, s) } {E}(\varepsilon x.(i(x),a) , \acute{g} \acute{t}h(g,t)) : 
{ C}}$

\hfill{$\triangleright_{mxl}  
\quad \displaystyle{  a: {A} \quad f(a) =_r i(a) : {B} (a)  
\atop  d(f/g , a/t ) =_s h(i/g , a/t):{ C}}$}

\smallskip

\noindent Associated rewritings:\\
$\mu (\xi_1 (r) , s) \triangleright_{mxr} s$\\
$\mu (\xi_2 (r) , s) \triangleright_{mxl} s$

\end{definition}


\begin{definition}[$\eta_{rewr}$]\ \\
$\eta_{rewr}$- $\times$-{\it reduction}

$\displaystyle{\displaystyle{ x=_r y : {A} \times {B} 
\over
{FST}(x) =_{\mu_1 ( r)} {FST}(y) : {A}} \times\mbox{\it 
-elim}  \  \displaystyle{ x=_r y : {A} \times {B} \over
{SND}(x) =_{\mu_2 ( r)} {SND}(y) : {B}
}\times\mbox{\it -elim} 
\over
\langle {FST}(x) , {SND}(x) \rangle =_{\xi ( \mu_1 ( r),\mu_2(r))}  \langle 
{FST} (y), {SND} (y) \rangle : {A} \times {B}
}\times\mbox{\it -intr} $

\hfill{$\displaystyle{\triangleright_{mx}} \   \displaystyle{x =_r y:  {A} \times {B}}$}

\smallskip

\noindent  $\eta_{rewr}$- $+$-{\it reduction}

\noindent $\displaystyle{\displaystyle{ \atop c=_t d : {A} +  
{B}}  \displaystyle{ [a_1 =_r a_2 : {A} ] \over {i}  (a_1) =_{\xi_1 (r)} {i} (a_2) : {A} +{B}}+\mbox{\it -intr} \displaystyle{ [ b_1 =_s b_2: {B}
 ] \over {j}(b_1) =_{\xi_2 (s)} {j}(b_2) : {A} + 
{B}}+\mbox{\it -intr} \over {D}(c, \acute{a_1}{i}(a_1), \acute{b_1}{j}(b_1)) =_{\mu ( t, \xi_1 (r) , \xi_2 (s))}
 {D}(d, \acute{ a_2}{i}(a_2), \acute{b_2} {j}(b_2))}+\mbox{\it -elim}$

\hfill{$\triangleright_{mxx} \quad c=_t d : {A} +{B}$}

\smallskip





\noindent $\Pi$-$\eta_{rewr}$-{\it reduction}

\noindent $\displaystyle{\displaystyle{[t:{A}] \quad c =_r d: \Pi x:{ A}.{B} (x) \over {APP}(c,t) =_{\nu (r)} {APP} (d,t): {
B}(t)} \Pi\mbox{\it -elim} \over \lambda t.{APP}(c,t) =_{\xi (\nu 
(r)) } \lambda t.{APP} (d,t) : 
\Pi t:{A}.{B}(t)}\Pi\mbox{\it -intr}$

\hfill{$\triangleright_{xmr} \qquad c=_r d: \Pi x:{A}.{B}(x)$}\\
where $c$ and $d$ do not depend on $x$.

\smallskip

\noindent $\Sigma$-$\eta_{rewr}$-{\it reduction}

\noindent $\displaystyle{\displaystyle{ \atop c=_sb: \Sigma x:{A}.{B} (x)} 
\quad  \displaystyle{ [t:{A}] \quad [g(t) =_r h(t) : {B}(t)] 
\over  \varepsilon y.(g(y),t) =_{\xi_2 (r) } \varepsilon y.(h(y), t) :  
\Sigma y :{A}.{B} (y)}\Sigma\mbox{\it -intr} \over {E} 
(c,  \acute{g}\acute{t}\varepsilon y.(g(y),t)) =_{\mu (s, \xi_2 (r))} 
{E} (b, \acute{h} \acute{t} \varepsilon y.(h(y) , t)) : 
\Sigma y :{A}.{B} (y)}\Sigma\mbox{\it -elim}$

\hfill{$\triangleright_{mxlr} \quad c =_s b : \Sigma x:{A}.{B} (x)$}

\smallskip

\smallskip\noindent
Associated rewritings:\\
$\xi (\mu_1 (r),\mu_2(r)) \triangleright_{mx} r$\\
$\mu (t, \xi_1 (r) , \xi_2 (s) ) \triangleright_{mxx} t$\\
$\xi (\nu (r)) \triangleright_{xmr} r$\\
$\mu ( s , \xi_2 (r)) \triangleright_{mxlr} s$
\end{definition}



\begin{definition}[$\sigma$ and $\tau$]\ \\
$\displaystyle{\displaystyle{x=_r y : {A} \quad y =_s w : {A}  
\over x =_{\tau(r,s)}  w : {A}} \over w=_{\sigma(\tau(r,s ))} x : {A}} \quad \triangleright_{stss} \quad 
\displaystyle{\displaystyle{  y=_s w : {A} \over w=_{\sigma(s)} y 
:  {A}} \quad \displaystyle{x=_r y : {A} \over y=_{\sigma(r)} 
x :  {A}} \over w=_{\tau(\sigma(s),\sigma(r))} x : {A}}$\\
\ \\
Associated rewriting:\\
$\sigma(\tau(r,s)) \triangleright_{stss} \tau(\sigma(s), \sigma(r))$

\end{definition}

\begin{definition}[$\sigma$ and ${\tt sub}$]\ \\
$$\displaystyle{\displaystyle{ x =_r {\cal C}[y] : {A} \quad y =_s w:  {A'} \over x=_{{\tt sub_L}(r,s)}{\cal C}[w] : {A}} \over 
{\cal C}[w] =_{\sigma( {\tt sub_L}(r,s))} x : {A}} \quad 
\triangleright_{ssbl}  \quad \displaystyle{\displaystyle{y=_s w : {A'} \over w=_{\sigma(s)} y : {A'}} \quad   
\displaystyle{x =_r {\cal C}[y]:{A} \over {\cal C}[y] =_{\sigma(r)} x : {A}} \over {\cal C}[w] =_{{\tt sub_R}(\sigma(s), 
\sigma(r))} x : {A}}$$

$$\displaystyle{\displaystyle{ x=_r y : {A'} \quad {\cal C}[y] =_s w:  {A} \over {\cal C}[x] = _{{\tt sub_R}(r,s)} w : {A}} \over 
w=_{ \sigma({\tt sub_R}(r,s))} {\cal C}[x] : {D}} \quad \triangleright_{ssbr} 
\quad  \displaystyle{\displaystyle{{\cal C}[y] =_s w : {A} \over w =_{\sigma(s)} {\cal C}[y] : 
{A} } \quad \displaystyle{x=_r y : {A'} \over y=_{\sigma(r)} x 
:  {A'}} \over w=_{{\tt sub_L}(\sigma(s),\sigma(r))} {\cal C}[x]  : {A}} $$
Associated rewritings:\\
$\sigma({\tt sub_L}(r,s)) \triangleright_{ssbl} {\tt sub_R}(\sigma(s),  \sigma(r))$ \\
$\sigma ({\tt sub_R} (r,s)) \triangleright_{ssbr} {\tt sub_L} ( \sigma (s) ,  \sigma (r))$

\end{definition}

\begin{definition}[$\sigma$ and $\xi$]\ \\
$$\displaystyle{\displaystyle{ x=_r y : {A} \over {i}(x) =_{\xi_1 
(r)}  {i}(y) : {A} +{B}} \over {i}(y) =_{\sigma(\xi_1  (r))} {i}(x) : {A} +{B}} \quad \triangleright_{sx} 
\quad  \displaystyle{\displaystyle{ x=_r y : 
{A} \over y =_{\sigma(r)} x : {A}} \over {i}(y) =_{\xi_1 
(  \sigma (r))} {i} (x) : {A} +{B}}$$

$$\displaystyle{\displaystyle{ x=_r y : {A} \quad z=_s w : {B} 
\over  \langle x,z \rangle =_{\xi(r,s)} \langle y,w \rangle : {A} 
\times   {B}} \over \langle y,w \rangle =_{\sigma(\xi( r,s))} 
\langle  x,z \rangle : {A} \times {B}} \quad \triangleright_{sxss} \quad  
\displaystyle{\displaystyle{x=_r y : {A} \over y=_{\sigma(r)} x :  {A}}  \quad \displaystyle{ z=_s w : {B} \over
w =_{\sigma(s)} z : {B}} \over \langle y,w \rangle  =_{\xi(\sigma(r),{\sigma(s))}} \langle x,z \rangle : {A} \times {B}}$$


$$\displaystyle{\displaystyle{\displaystyle{ [ x : {A} ] \atop f(x) =_s 
g(x)  : {B} (x)} \over \lambda x.f(x) =_{\xi (s)} \lambda x.g(x) :  
\Pi x:{A}.{B} (x)} \over   \lambda x.g(x) =_{\sigma (\xi (s))} \lambda x.f(x) : \Pi x:{A}.{B} (x)}  \  
\triangleright_{smss}  \  \displaystyle{\displaystyle{\displaystyle{ [ x : {A} 
]  \atop f(x) =_s g(x) : {B} (x)} \over  g(x) =_{\sigma(s)} f(x) 
:  {B} (x)} \over \lambda x.g(x) =_{\xi ( \sigma (s))} \lambda x.f(x) 
: \Pi x:{A}.{B} (x)} $$

\smallskip

\noindent Associated rewritings:\\
$\sigma(\xi (r)) \triangleright_{sx} \xi ( \sigma(r))$\\
$\sigma(\xi (r, s)) \triangleright_{sxss} \xi ( \sigma(r), \sigma(s))$\\
$\sigma(\xi (s) \triangleright_{smss} \xi ( \sigma(s))$\\

\end{definition}

\begin{definition}[$\sigma$ and  $\mu$]\ \\
$$\displaystyle{\displaystyle{ x =_r y : {A} \times {B} \over {FST}(x) =_{\mu_1 (r)} {FST} (y) : {A}} \over {FST}(y) =_{\sigma (\mu_1 (r))} {FST} (x) : {A}} \quad
 \triangleright_{sm} \quad   
\displaystyle{\displaystyle{ x=_r y : {A} \times {B} \over  
y=_{\sigma(r)} x : {A} \times {B}} \over {FST} (y) =_{\mu_1 
(\sigma (r))} {FST}(x) : {A}}$$

\noindent $\displaystyle{\displaystyle{ x=_s y : {A} \quad f=_r g : {A} 
\to  {B} \over {APP} (f,x) =_{\mu (s,r)} {APP }(g,y) : {B}}  \over {APP} (g,y) =_{\sigma(\mu (s,r))} {APP}(f,x) : {B}}$

\hfill{$\triangleright_{smss} \quad \displaystyle{\displaystyle{  x=_s y : {A} 
\over   y=_{\sigma (s)}  x : {A}} \quad \displaystyle{ f=_r g : 
{A } \to {B} \over g=_{\sigma(r)} f : {A} \to {B}} 
\over  {APP}(g,y) =_{\mu (\sigma(s), \sigma(r))} {APP} 
(f,x)  : {B} }$}

\smallskip

\noindent $\displaystyle{\displaystyle{\displaystyle{\displaystyle{ \atop } 
\atop    x=_r y : {A} +{B}} \quad 
\displaystyle{\displaystyle{   [s: {A} ] \atop \vdots} \atop  d(s) =_u 
f(s)  : {C} } \quad \displaystyle{\displaystyle{ [t:{B}] \atop
\vdots } \atop  e(t) =_v g(t) : {C}} \over {D} (x, \acute{s}d(s),  \acute{t} e(t)) =_{\mu (r,u,v)} {D}(y, \acute{s}f(s),  \acute{t}g(t)) : {C}} \over {D} (y, \acute{s}f(s),  \acute{t}g(t)) : {C} =_{\sigma(\mu (r,u,v))} {D}  (x, \acute{s}d(s), \acute{t}e(t)) : {C}}$

\hfill{$\triangleright_{smsss} \displaystyle{\displaystyle{\displaystyle{ \atop x=_r y : {A} +  
{B}} \over y=_{\sigma (r)} x : {A} +{B}} \quad  
\displaystyle{\displaystyle{[s:{A} ] \atop d(s) =_u f(s) : {C}} 
\over  f(s) =_{\sigma(u)} d(s) : {C}} \quad  
\displaystyle{\displaystyle{ [t:{B}] \atop e(t) =_v g(t) : {C} }  
\over g(t) =_{\sigma (v)} e(t) : {C}} \over {D}  
(y,\acute{s}f(s), \acute{t}g(t)) =_{\mu (\sigma(r), \sigma
(u),  \sigma (v))} {D} (x, \acute{s}d(s), \acute{t} e(t)) 
:  {C}}$}

\smallskip

\noindent $\displaystyle{\displaystyle{\displaystyle{ \atop e=_s b : \Sigma x:{A}.{B} (x)} \quad \displaystyle{ [t: {A}, \; g(t) : {B} (t) 
]  \atop d(g,t) =_r f(g,t) : {C}} \over {E} (e,\acute{g}\acute{t} d(g,t)) =_{\mu (s,r)} {E} (b, \acute{g} \acute{t} f(g,t)) :  
{C}} \over {E} (b,\acute{g} \acute{t} f(g,t)) =_{\sigma
(\mu (s,r))} {E} (e, \acute{g} \acute{t} d(g,t)) : {C}}$

\hfill{$ \triangleright_{smss} \displaystyle{\displaystyle{ \displaystyle{ \atop e=_s b : \Sigma 
x:{A}.{B} (x)} \over b=_{\sigma(s)} e : \Sigma x:{A}.{B} (x)}\quad \displaystyle{\displaystyle{ [t: {A}, \; g(t) :  
{B} (t) ] \atop d(g,t) =_r f(g,t) : {C}} \over f(g,t)  =_{\sigma(r)} d(g,t) : {C}} \over {E} (b,\acute{g} \acute{t} f(g,t))  =_{\mu (\sigma (s), \sigma (r))} {E} (e, \acute{g} \acute{t}d(g,t)) : {C}}  $}

\smallskip\noindent
Associated rewritings:\\
$\sigma(\mu (r)) \triangleright_{sm} \mu ( \sigma(r))$\\
$\sigma(\mu (s, r)) \triangleright_{smss} \mu ( \sigma(s), \sigma(r))$\\
$\sigma(\mu (r, u,v)) \triangleright_{smsss} \mu ( \sigma(r),  \sigma(u),\sigma(v))$

\end{definition}


\begin{definition}[$\tau$ and ${\tt sub}$]\ \\
$\displaystyle{\displaystyle{ x=_r {\cal C}[y]:A \quad y =_s w: {A'} \over x =_{{\tt sub_L} (r,s)} {\cal C}[w]: {A}} \quad
\displaystyle{ \atop {\cal C}[w] =_t z : { A}} \over x =_{ \tau  ({\tt sub_L} (r,s) , t)} z : {A}}$

\hfill{$  \triangleright_{tsbll} \ \  \displaystyle{\displaystyle{  \atop  
x=_r {\cal C}[y]: {A}} \quad  \displaystyle{ y =_s w : {A'}  
\quad {\cal C}[w] =_t z : {A} \over {\cal C}[y] =_{{\tt 
sub_R}(s,t)}  z : {A}} \over x=_{\tau (r, {\tt sub_R}(s,t))} z 
:  {A}}$}

\smallskip

\noindent $\displaystyle{\displaystyle{ y=_s w: {A} \quad {\cal C}[w] =_t z 
:  {A} \over {\cal C}[y] =_{{\tt sub_R} ( s,t)} z : { A}} \quad
\displaystyle{ \atop z =_u v : { A}} \over {\cal C}[y] =_{ \tau  ({\tt sub_R} (s,t) , u)} v : {A}}$

\hfill{$ \triangleright_{tsbrl} \   \displaystyle{ \displaystyle{ \atop y =_s w : {D'}} \ \  
\displaystyle{{\cal C}[w] =_t z : {A} \quad z =_u v : {A} \over 
{\cal C}[w] =_{\tau(t,u)} v : {A}} \over {\cal C}[y]=_{{\tt sub_R } ( s, \tau ( t,u))} v : {A}}$}

\smallskip

\noindent $\displaystyle{\displaystyle{ \atop x=_r {\cal C}[z] : {A}} \quad
\displaystyle{{\cal C}[z] =_{\rho} {\cal C}[z] : {A} 
\quad  z =_s w : {A'} \over {\cal C}[z] =_{{\tt sub_L} (\rho , s)} {\cal C}[w] : {A}} \over x =_{\tau ( r, {\tt 
sub_L}  (\rho, s))} {\cal C}[w] : {A}}$

\hfill{$ \triangleright_{tsblr} \   \displaystyle{ x=_r {\cal C}[z] : {A}  \quad z =_s w : {A'}  \over x =_{{\tt sub_L} (r,s)} {\cal C}[w]: {A}}$}

\smallskip

$\displaystyle{\displaystyle{ \atop x =_r {\cal C}[w]: {A}}  \quad
\displaystyle{ w=_s z : {A'} \quad {\cal C}[z]=_{\rho} {\cal C}[z] : {A} \over {\cal C}[w] =_{{\tt sub_R} (s, \rho)} {\cal C}[z] : {A}} \over x =_{\tau(r, {\tt sub_R}  
(s, \rho))} {\cal C}[z] : {A}}$

\hfill{$ \triangleright_{tsbrr} \   \displaystyle{ x =_r {\cal C}[w] : {D} \quad w =_s z : {A'}  
\over x =_{{\tt sub_L} (r,s)} {\cal C}[z] : {A}}$}

\end{definition}

\begin{definition}[$\tau$ and $\tau$]\ \\
$\displaystyle{\displaystyle{ x=_t y:A \quad y =_r w: {A} \over x =_{\tau(t,r)}w: {A}} \quad
\displaystyle{ \atop w =_s z : {A}} \over x =_{\tau (\tau (t,r) , s)} z : {A}}$

\hfill{$\triangleright_{tt} \quad  
\displaystyle{\displaystyle{ \atop x =_t y : {A}} \quad  
\displaystyle{ y=_r w : {A} \quad w=_s z : {A} \over  y=_{\tau(r,s)} z : {A}} \over x=_{\tau(t,\tau(r,s))} z :  
{A}}$}

\smallskip\noindent
Associated rewritings: \\
$\tau({\tt sub_L}(r,s),t) \triangleright_{tsbll} \tau (r, {\tt sub_R}(s,t))$\\
$\tau ({\tt sub_R } (s,t), u)) \triangleright_{tsbrl} {\tt sub_R} ( s, \tau  (t,u))$\\
$\tau (r, {\tt sub_L} (\tau , s)) \triangleright_{tsblr} {\tt sub_L} (r,s)$\\
$\tau (r, {\tt sub_R} (s, \tau )) \triangleright_{tsbrr} {\tt sub_L} (r,s)$\\
$\tau(\tau (t,r),s) \triangleright_{tt} \tau(t,\tau(r,s))$

\end{definition}

By analysing all cases of redundant proofs involving equality we arrive at
following set of associated rewriting rules. (NB. In the same way the definitional equalities (coming from rewriting rules) over terms of 
the $\lambda$ calculus had to be given names -- $\beta$, $\eta$, $\xi$, $\mu$, etc. --, we will need to assign a name to each rewriting rule
for terms representing computational paths. For the lack of a better naming choice at this point, we have tried to use abbreviations related
to the operations involved.)
\begin{definition}[$LND_{EQ}$-$TRS$]\label{rewriting-system}\ \\
1. $\sigma(\rho) \triangleright_{sr} \rho$ \\ 
2. $\sigma(\sigma(r)) \triangleright_{ss} r$\\ 
3. $\tau({\cal C}[r] , {\cal C}[\sigma(r)]) \triangleright_{tr}  {\cal C }[\rho]$\\ 
4. $\tau({\cal C}[\sigma(r)], {\cal C}[r]) \triangleright_{tsr} {\cal C}[\rho]$\\ 
5. $\tau({\cal C}[r], {\cal C}[\rho]) \triangleright_{rrr} {\cal C}[r]$\\ 
6. $\tau({\cal C}[\rho], {\cal C}[r]) \triangleright_{lrr} {\cal C}[r]$ \\ 
7. ${\tt sub_L}({\cal C}[r], {\cal C}[\rho]) \triangleright_{slr} {\cal C}[r]$\\ 
8. ${\tt sub_R}({\cal C}[\rho], {\cal C}[r]) \triangleright_{srr} {\cal C}[r]$ \\
9. ${\tt sub_L} ({\tt sub_L} (s, {\cal C}[r]), {\cal C}[\sigma(r)]) \triangleright_{sls} s$\\
10. ${\tt sub_L} ( {\tt sub_L} (s , {\cal C}[\sigma(r)]) , {\cal C}[r]) \triangleright_{slss} s$\\ 
11. ${\tt sub_R} ({\cal C}[s], {\tt sub_R} ({\cal C}[\sigma(s)],r)) \triangleright_{srs} r$\\ 
12. ${\tt sub_R} ({\cal C}[\sigma(s)], {\tt sub_R} ({\cal C}[s] ,  r )) \triangleright_{srrr} r$\\ 
13. 
 $\mu_1 ( \xi_1 ( r))\triangleright_{mx2l1} r$\\
14. $\mu_1 ( \xi_\land ( r,s))\triangleright_{mx2l2} r$\\
15.
 $\mu_2 ( \xi_\land ( r,s))\triangleright_{mx2r1} s$\\
 16.
$\mu_2 ( \xi_2 ( s))\triangleright_{mx2r2} s$\\
17. 
$\mu ( \xi_1 (r) , s , u) \triangleright_{mx3l} s$\\ 
18. 
 $\mu (\xi_2 (r) , s , u) \triangleright_{mx3r} u$\\ 
19.
$\nu (\xi (r)) \triangleright_{mxl} r$\\ 
20.
$\mu (\xi_2 (r) , s) \triangleright_{mxr} s$\\ 
21.
$\xi ( \mu_1 (r),\mu_2(r) ) \triangleright_{mx} r$ \\ 
22.
$\mu ( t, \xi_1 (r), \xi_2 (s)) \triangleright_{mxx} t$ \\ 
23. 
$\xi ( \nu (r) ) \triangleright_{xmr} r$ \\ 
24. 
$\mu (s,\xi_2 (r)) \triangleright_{mx1r} s$\\ 
25. $\sigma(\tau(r,s)) \triangleright_{stss} \tau(\sigma(s),  \sigma(r))$\\ 
26. $\sigma({\tt sub_L}(r,s)) \triangleright_{ssbl} {\tt sub_R}(\sigma(s), \sigma(r))$\\ 
27. $\sigma ({\tt sub_R} (r,s)) \triangleright_{ssbr} {\tt sub_L} (\sigma
(s),  \sigma (r))$\\ 
28. $\sigma(\xi (r)) \triangleright_{sx} \xi ( \sigma(r))$\\ 
29. $\sigma(\xi (s, r)) \triangleright_{sxss} \xi ( \sigma(s),  \sigma(r))$\\ 
30. $\sigma(\mu (r)) \triangleright_{sm} \mu ( \sigma(r))$\\ 
31. $\sigma(\mu (s, r)) \triangleright_{smss} \mu (\sigma(s),  \sigma(r))$\\ 
32. $\sigma(\mu (r,u,v)) \triangleright_{smsss} \mu ( \sigma(r),\sigma(u),\sigma(v))$\\
33. $\tau (r, {\tt sub_L} (\rho , s)) \triangleright_{tsbll} {\tt sub_L}  (r,s)$\\ 
34. $\tau (r, {\tt sub_R} (s, \rho)) \triangleright_{tsbrl}  {\tt 
sub_L} (r,s)$\\ 
35. $\tau({\tt sub_L}(r,s),t) \triangleright_{tsblr} \tau (r, {\tt 
sub_R} (s,t))$\\ 
36. $\tau ({\tt sub_R} (s,t),u) \triangleright_{tsbrr} {\tt sub_R} (s, \tau  (t,u))$\\ 
37. $\tau(\tau(t,r),s) \triangleright_{tt} \tau(t,\tau (r,s)) $\\
38. $\tau ({\cal C}[u], \tau ({\cal C}[\sigma(u)] , v)) \triangleright_{tts} v$\\
39. $\tau ({\cal C}[\sigma(u)] , \tau ({\cal C}[u] , v)) \triangleright_{tst} u$.
\end{definition}

\subsection{Termination property for the $LND_{EQ}$-$TRS$}

\begin{theorem}[Termination property for $LND_{EQ}$-$TRS$]
$LND_{EQ}$-$TRS$ is terminating.
\end{theorem}

The proof of the termination property for $LND_{EQ}$-$TRS$ is obtained by 
using a special kind of ordering: {\em recursive path ordering\/}, proposed 
by N.\ Dershowitz in 1982 \cite{dershowitz82}:

\begin{definition}[recursive path ordering]
Let $>$ be a partial ordering 
on a set of operators F. The recursive path ordering $>^*$ on the set 
T(F) of terms over F is defined recursively as follows:

$$ s = f(s_1,\ldots , s_m) >^* g(t_1,\ldots, t_n) = t,$$
if and only if
\begin{enumerate}
\item $f=g$ and $\{ s_1, \ldots , s_m\} \gg^*  \{ t_1, \ldots , t_n\}$, or
\item $f>g$ and $\{s\} \gg^* \{t_1, \ldots , t_n\}$, or
\item $f \ngeq g$ and $\{s_1, \ldots , s_m \} \gg^* $ or $=$ $\{t\}$
\end{enumerate}
where $\gg^*$ is the extension of $>^*$ to multisets.

\end{definition}

Note that this definition uses the notion of ordering on multisets. A 
given partial ordering $>$ on a set $S$ may be extended to a partial 
ordering $\gg$ on finite multisets of elements of $S$, wherein a multiset 
is reduced by removing one or more elements and replacing them with any 
finite number of elements, each oh which is smaller than one of the 
elements removed \cite{dershowitz82}.

The proof of termination property via a recursive path ordering is made 
by showing that for all rules $e \to d$ of the system, $e >^* d$.

The {\em recursive path ordering\/} can be extended in order to allow some
function of a term $f(t_1,\ldots,t_n)$ to play the role of the operator $f$.
As explained in \cite{dershowitz82}, we can consider the $k$-th operand $t_k$ to
be the operator, and compare two terms by first recursively comparing their
$k$-th operands.

In the proof of termination property for the 
$LND_{EQ}$-$TRS$, we use 
the precedence ordering on the rewrite operators  for the rules from $1$ to $32$
defined as follows:
$$\begin{array}{l}
\sigma > \tau > \rho, \\
\sigma > \xi, \\
\sigma > \xi_\land, \\
\sigma > \xi_1, \\
\sigma > \xi_2, \\
\sigma > \mu, \\
\sigma > \mu_1, \\
\sigma > \mu_2, \\
\sigma > {\tt sub_L}, \\
\sigma > {\tt sub_R}, \\
\tau > {\tt sub_L}
\end{array}$$

We can combine the recursive path idea used for the rules 1--34 with an
extension of recursive path ordering for rules from 35 to 37, where the first
operand is used as operator. When comparing two ``$\tau$'' and
``$\tau$ with ${\tt sub_R}$'', we use the first operand as operator.
This proof is similar to Example (H) given in pp.\ 299--300 of \cite{dershowitz82}.
 
The confluence proof is built by the Knuth--Bendix superposition algorithm 
applied to the rules of the system.

We have proved termination and confluence of the rewriting system
$LND_{EQ}$-$TRS$  \cite{teseju,deOliveira2,deOliveira3}. As a matter of fact, rules 38 and 39 came out of the Knuth--Bendix completion procedure applied to the
the rewriting system.
As we have previously pointed out, although the rewriting system is terminating and confluent, we have observed an interesting
phenomenon here: there may be more than one normal proof of an equality statement. This is not a contradiction since the confluence property 
only says that the term for the equality reason can be brought to a unique normal form regardless of the order in which it is reduced. But there may
be a different, yet normal/canonical, proof of the same equality statement.

\section{Conclusion}
Motivated by looking at equalities in type theory as arising from the existence of computational paths between two 
formal objects, our purpose here was to offer an alternative perspective (to the one prevailing on the literature) on the role and the power of the 
so-called identity types, as well as of the notion of propositional equality as formalised in the so-called Curry-Howard functional interpretation. 
We started by recalling our previous observation \cite{absldsequality} pertaining to 
the fact that the formulation of the identity type by Martin-L\"of, both in the intensional and in the extensional versions, did not take
into account an important entity, namely, identifiers for sequences of rewrites, and this has led to a false dichotomy.

Next, by considering as sequences of rewrites and substitution, we have shown that it comes a rather natural fact that
two (or more) distinct proofs may be yet canonical and are none to be preferred over one another. By looking at proofs of equality as rewriting 
(or computational) paths this approach fits well with the recently proposed connections between type theory and homotopy theory via identity types, 
since elements of identity types will be, concretely, paths (or homotopies). In the end, our formulation
of a proof theory for propositional equality is still very much in the style of type-theoretic identity
types which, besides being a reformulation of Martin-L\"of's own intensional identity types into one which dissolves what we see as a false dichotomy, 
turned out to validate the groupoid laws as uncovered by Hofmann \& Streicher as well as to refute the principle of uniqueness of identity proofs.

\small

\bibliographystyle{plain}

\appendix

\section*{Appendix: Definitional Equality Rules}
$$\begin{array}{llll}
(\beta) & \displaystyle{\displaystyle{{\ \atop {a:A}}\quad \displaystyle{{[x:A]} \atop {b:B}}}\over (\lambda x.b)a =_\beta b[a/x]:B}\\
&\\
(\eta) & \displaystyle{{f:(\Pi x:A)B}\over \lambda x.{APP}(f,x) =_\eta f:(\Pi x:A) B} \ (x\notin FV(f))\\
&\\
(\rho) & \displaystyle{a:A\over a=_\rho a:A}\\
&\\
(\mu) & \displaystyle{{a=_s a':A\quad f:(\Pi x:A)B}\over {{APP}(f,a)=_{\mu(s)}{APP}(f,a'):B}}\qquad & (\tau) & \displaystyle{{a=_s a':A \qquad a'=_t a'':A} \over {a=_{\tau(s,t)}a'':A}}\\
&\\
(\nu) & \displaystyle{{a:A\quad f=_s f':(\Pi x:A)B}\over {{APP}(f,a)=_{\nu(s)}{APP}(f',a):B}}\qquad & (\sigma) & \displaystyle{{a=_s a':A} \over {a'=_{\sigma(s)}a:A}} \\
&\\
(\xi) & \displaystyle{\displaystyle{{[x:A]} \atop {f(x)=_s f'(x):B}}\over {\lambda x.f(x)=_{\xi(s)}\lambda x.f'(x):(\Pi x:A)B}}\\
&\\
(\mu_1) & \displaystyle{{p=_s q:A\times B} \over {{FST}(p)=_{\mu_1(s)}{FST}(q):A}}  & (\mu_2) & \displaystyle{{p=_s q:A\times B} \over {{SND}(p)=_{\mu_2(s)}{SND}(q):B}}   \\
&\\
(\xi_\land) & \displaystyle{{a=_r a':A \quad b=_sb':B} \over {\langle a,b\rangle=_{\xi_\land(r,s)}\langle a',b'\rangle:A\times B}}  \\
&\\
(\xi_1) & \displaystyle{{a=_s a':A \quad b:B} \over {\langle a,b\rangle=_{\xi_1(s)}\langle a',b\rangle:A\times B}}  & (\xi_2) & \displaystyle{{a:A\quad  b=_sb':B} \over {\langle a,b\rangle=_{\xi_2(s)}\langle a,b'\rangle:A\times B}}\\
&\\
(\xi_1) & \displaystyle{{a=_s a':A} \over {i (a)=_{\xi_1(s)}i(a'):A+B}}  & (\xi_2) & \displaystyle{{b=_s b':B} \over {j(b)=_{\xi_2(s)}j(b'):A+B}} \\
&\\
(\nu) & \displaystyle{{\displaystyle{\ \atop {p=_sq: A+B}}\quad \displaystyle{{[x:A]} \atop {f(x):C}}\quad \displaystyle{{[y:B]} \atop {g(y):C}}}\over {\displaystyle{D(p,\acute{x}f(x),\acute{y}g(y))=_{\nu(s)}D(q,\acute{x}f(x),\acute{y}g(y)):C}}}\\
&\\
(\mu_1) & \displaystyle{{\displaystyle{\ \atop {p: A+B}}\quad \displaystyle{{[x:A]} \atop {f(x)=_sf'(x):C}}\quad \displaystyle{{[y:B]} \atop {g(y):C}}}\over {\displaystyle{D(p,\acute{x}f(x),\acute{y}g(y))=_{\mu_1(s)}D(p,\acute{x}f'(x),\acute{y}g(y)):C}}}\\
&\\
(\mu_2) & \displaystyle{{\displaystyle{\ \atop {p: A+B}}\quad \displaystyle{{[x:A]} \atop {f(x):C}}\quad \displaystyle{{[y:B]} \atop {g(y)=_sg'(y):C}}}\over {\displaystyle{D(p,\acute{x}f(x),\acute{y}g(y))=_{\mu_2(s)}D(p,\acute{x}f(x),\acute{y}g'(y)):C}}}\\
&\\
(\xi_1) & \displaystyle{{a=_sa':A \quad b(a):B(a)} \over {\varepsilon x.(b(x),a)=_{\xi_1(s)}\varepsilon x.(b(x),a'):(\Sigma x:A)B(x)}}  & \\
&\\
(\xi_2) & \displaystyle{{a:A\quad  b(a)=_sb'(a):B(a)} \over {\varepsilon x.(b(x),a)=_{\xi_2(s)}\varepsilon x.(b'(x),a):(\Sigma x:A)B(x)}}\\
&\\
(\nu) & \displaystyle{{\displaystyle{\ \atop {p=_sq: (\Sigma x:A)B(x)}}\quad \displaystyle{{[t:A,\ f(t):B(t)]} \atop {h(t,f):C}}}}\over {\displaystyle{E(p,\acute{f}\acute{t}h(t,f))=_{\nu(s)}E(q,\acute{f}\acute{t}h(t,f)):C}}&
(\mu) & \displaystyle{{\displaystyle{\ \atop {p: (\Sigma x:A)B(x)}}\quad \displaystyle{{[t:A,\ f(t):B(t)]} \atop {h(t,f)=_s h' (t,f):C}}}}\over {\displaystyle{E(p,\acute{f}\acute{t}h(t,f))=_{\mu(s)}E(p,\acute{f}\acute{t}h' (t,f)):C}}

\end{array}$$

\end{document}